\documentclass[fleqn,usenatbib]{mnras}
\usepackage{newtxtext,newtxmath}
\usepackage[T1]{fontenc}
\usepackage{ae,aecompl}
\usepackage{graphicx}	
\usepackage{amsmath}	
\usepackage{amssymb}	

\hypersetup{pdfauthor={N.P.M. Kuin},
               pdftitle={Swift spectra of AT2018cow: A White Dwarf Tidal Disruption Event},
               pdfkeywords={TDE, stars:black holes, stars:white dwarfs},
               bookmarksnumbered=true}

\def\Td{$T_{\rm d}$}
\def\NH{$N_{\rm H}$}
\def\Swift{{\em Swift} }
\def\revv{} 
\def\rev2{}

\title[AT2018cow: A WD TDE]{ {\em Swift} spectra of AT2018cow: A White Dwarf Tidal Disruption Event?}

\author[N. P. M. Kuin et al.]{N.~Paul~M. Kuin,$^1$\thanks{email: n.kuin@ucl.ac.uk }
Kinwah Wu,$^1$
Samantha Oates,$^{2}$
Amy Lien,$^{3,13}$
Sam Emery,$^1$
\newauthor
Jamie A. Kennea,$^6$
Massimiliano de Pasquale,$^4$
Qin Han,$^1$
Peter J. Brown,$^{5,9}$ 
\newauthor
Aaron Tohuvavohu,$^6$
Alice Breeveld,$^1$
David N. Burrows,$^6$
S. Bradley Cenko,$^{7,12}$
\newauthor
Sergio Campana,$^{8}$
Andrew Levan,$^2$
Craig Markwardt,$^{7}$,
Julian P. Osborne,$^{10}$
\newauthor
Mat J. Page,$^1$
Kim L. Page,$^{10}$
Boris Sbarufatti,$^6$ 
Michael Siegel,$^6$ and
\newauthor
Eleonora Troja$^{7,11}$
\\
$^1$Mullard Space Science Laboratory, University College London, Holmbury St.~Mary, Dorking, Surrey RH5 6NT, UK\\
$^2$Department of Physics. University of Warwick, Coventry, CV4 7AL, UK \\ 
$^3$Department of Physics, University of Maryland, Baltimore County, 1000 Hilltop Circle, Baltimore, MD 21250, USA \\ 
$^4$Department of Astronomy and Space Sciences, Istanbul University, Beyazit\it, 34119, Istanbul, Turkey\\ 
$^5${Department of Physics and Astronomy,
Texas A\&M University, 4242 TAMU, College Station, TX 77843, USA }\\
$^6$Department of Astronomy and Astrophysics, Pennsylvania State University. 
  525 Davey Laboratory, University Park, PA 16802, USA\\ 
$^7$Astrophysics Science Division, NASA Goddard Space Flight Center, 8800 Greenbelt Road, Greenbelt, MD 20771, USA \\ 
$^8$INAF-Osservatorio Astronomico di Brera, via E. Bianchi 46, 23807, Merate, Italy \\ 
$^9${George P. and Cynthia Woods Mitchell Institute for Fundamental Physics \& Astronomy} \\ 
$^{10}$Department of Physics and Astronomy, University of Leicester, University Road, Leicester LE1 7RH, UK. \\
$^{11}$ Department of Astronomy, University of Maryland, College Park, MD 20742-4111, USA\\
$^{12}$ Joint Space-Science Institute, University of Maryland, College Park, MD 20742, USA\\
$^{13}$Center for Research and Exploration in Space Science and Technology (CRESST) and NASA Goddard Space Flight Center, \\
\ \ \ \   Greenbelt, MD 20771, USA\\
}

\date{Accepted by MNRAS 28 December 2018. preprint}

\pubyear{2019}
   \volume{{\rm 000}}
\begin{document}
\label{firstpage}
\pagerange{\pageref{firstpage}--\pageref{lastpage}}
\maketitle

\begin{abstract}
The bright transient AT2018cow has been unlike any other known type of transient. 
Its high brightness, rapid rise and decay and initially nearly featureless spectrum are unprecedented and difficult to explain using models for similar burst sources.
We present evidence for faint $\gamma$-ray emission continuing for at least 8 days, and featureless spectra in the ultraviolet bands --  both unusual for eruptive sources. 
The X-ray variability of the source has a burst-like character. 
The UV-optical spectrum does not show any CNO line but is well described by a blackbody.
We demonstrate that a model invoking the tidal disruption of a $0.1 - 0.4\;M_\odot$ Helium White Dwarf\,(WD) by a $10^5-10^6\;M_\odot$  Black Hole\,(BH) {\revv located in the outskirts of galaxy {\tt Z~137-068}} could provide an explanation for most of the characteristics shown in the multi-wavelength observations.  
A blackbody-like emission is emitted from  an opaque photosphere, formed by the debris of the WD disruption. 
Broad features showing up in the optical/infrared spectra in the early stage are probably velocity broadened lines produced in a transient high-velocity outward moving cocoon.
The asymmetric optical/infrared lines that appeared at a later stage  are emission from an atmospheric layer when it detached from thermal equilibrium with the photosphere, which undergoes more rapid cooling.   
The photosphere shrinks when its temperature drops, and the subsequent infall of the atmosphere produced asymmetric line profiles.
Additionally, a non-thermal jet might be present, emitting X-rays in the $10-150$\,keV band.

\end{abstract}




\section{Introduction}

The transient AT2018cow/ATLAS18qqn/SN2018cow was discovered at an offset of 6\arcsec (1.7\,kpc) from galaxy {\tt Z~137-068} 
\citep{2018ATel11742....1S} by the ATLAS wide-field survey \citep{2018TNSTR.838....1T} on 2018-06-16 10:35:38 UT (MJD 58285.44141, 
 referred to in this paper as the discovery date \Td) 
 at an AB magnitude $o$ = ${14.74\pm0.10}$\,mag 
 (the $o$-band covers $560-820\;{\rm nm}$,\footnote{http://www.fallingstar.com/specifications}).
A previous observation by \cite{2018ATel11738....1F} on MJD~58282.172  (3.3\,d before the discovery date) with the Palomar 48-inch in the $i$-band did not detect a source down to a limiting magnitude of $ i > 19.5$\,mag, while on MJD~58286 (\Td+0.75\,d) 
$i = 14.32\pm0.01$\,mag, nearly 5 magnitudes brighter - a rapid rise.
Maximum light occurred at MJD~58286.9  \citep[\Td+1.46\,d,][]{2018arXiv180705965P}).

Spectroscopic follow-up by \citet{2018ATel11732....1P}, and \citet{2018ATel11776....1P}  using the SPRAT on the Liverpool Telescope (402-800\,nm, with 2\,nm resolution) on MJD 58287.951 (\Td+1.56\,d) found a smooth spectrum.  \citet{2018ATel11736....1J} 
 reported the Ca\,II~H and K absorption lines close to  the redshift of the co-located galaxy, 
  proving that the transient was near that galaxy. 
Spectra taken on the Xinglong 2.16-m Telescope using the BFOSC showed weak broad bumps or dips in the spectrum \cite[][]{2018ATel11740....1X, 2018ATel11753....1I} 
{\revv which may be interpreted as highly velocity-broadened lines though Perley et at (2018a) considered the features as an absorption trough.} 
The velocity derived from the broadening of the presumably He emission 
  was $\approx 1.6\times 10^4\,{\rm km\,s}^{-1}$ on  \citep[\Td+4.1\,d,][]{2018arXiv180705965P}.
Intrinsic optical polarization was measured on days \Td+4.9 and 5.9\,d by \citet{2018ATel11789....1S}. 

At high energies the transient was detected by the {\em Neil Gehrels Swift Observatory} \citep[hereafter {\em Swift},][]{2004ApJ...611.1005G} XRT \citep{2005SSRv..120..165B} in the $0.3-10$\,keV band  \citep{2018arXiv180706369R},
NICER   \citep[$0.5-10$\,keV]{2018ATel11773....1M}, NuSTAR \citep[$3-60$\,keV]{2018ATel11775....1M}, 
  and INTEGRAL IBIS/SGRI \citep[$30-100$\,keV]{2018ATel11788....1F}.
A search for impulsive emission by Fermi/GBM \citep[$10-1000$\,keV]{2018ATel11793....1D}, Fermi/LAT \citep[$\geq 100$\,MeV]{2018ATel11808....1K}, the INSIGHT HXMT/HE \citep[$80-800$\,keV]{2018ATel11799....1H} and Astrosat CZTI \citep[$20-200$\,keV]{2018ATel11809....1S} was unsuccessful. 

In the radio a search of pre-outburst data by \cite{2018ATel11744....1D} found $3\sigma$ upper limits of $370\,\mu$Jy at 3\,GHz and $410\,\mu$Jy at 1.4\,GHz.
The transient was detected at 90 and 150\,GHz on \Td+4.5\,d with a flux density of $\approx$ 6\,mJy at 90 GHz \citep{2018ATel11749....1D}, at 350\,GHz on day 5.8 with flux density of $30.2\pm1.8$ mJy/beam \citep{2018ATel11781....1S} and with a $5\sigma$ detection at 15.5\,GHz of 0.5 mJy on  \Td+6.3\,d  \citep{2018ATel11774....1B}.
Further detections were reported on days \Td+10 and 11\,d at 9\,GHz and 34\,GHz, and on \Td+12 also at 5.5\,GHZ \citep{2018ATel11795....1D,2018ATel11818....1D}.

We will adopt a distance to the transient consistent with it being associated with the nearby galaxy {\tt Z\,137-068},
  which has a red-shift $z = 0.01414\pm0.00013$\footnote{NED refcode  2007SDSS6.C...0000}
The reddening towards the galaxy is low, $E(B-V) = 0.077$ \citep{2011ApJ...737..103S} 
  and \NH$_{\rm galactic} = 6.57\times 10^{20}\,{\rm cm}^{-2}$ \cite[][]{2013MNRAS.431..394Willingdale}. 
Adopting cosmological parameters $H_0 = 71.0\;{\rm km~s}^{-1}{\rm Mpc}^{-1}$, $\Omega_{\rm m}=0.27$, $\Omega_v=0.73$,   
\cite[][]{2011ApJS..192...14J} 
  the distance is $60\pm 4$\,Mpc\footnote{using NED/IPAC}. 

During our studies and preparation of this paper three other studies were published in preprint form \citep{2018arXiv180705965P,2018arXiv180706369R,2018arXiv180800969Perley}, and we discuss and use their results in our discussion of the nature of the transient whilst extending their analysis. {\revv As in this paper, \citet{2018arXiv180800969Perley} proposed that the transient could be a TDE and they discussed constraints on the TDE properties using recent models. 
In two further papers a more general analysis was made in terms of a central engine \citep[][]{2018arXiv181010880Ho,2018arXiv181010720M_Raf}, leaving open the nature of the source. }

We discuss our observations, and present a model derived from the observations in terms of the tidal disruption of a He white dwarf  by a  non-stellar mass black hole, i.e. a TDE-WD event, where the debris forms a photosphere which produces blackbody-like emission in the UV-optical bands and  with emission lines formed above the photosphere.  
Moreover, a rapidly expanding cocoon has become detached from the photosphere and envelops the system initially. 
It produces very broad emission features attributed to velocity broadened lines, i.e., the bumps seen by \cite{2018arXiv180705965P} {\revv and \cite{2018arXiv180800969Perley}}. 
Finally, a jet is associated with the event, responsible for the high-energy $\gamma$-ray and X-ray emission.

\section{Observations}

{\em Swift} started pointed observations of AT2018cow with all three instruments: the Burst Alert Telescope \citep[BAT]{2005SSRv..120..143B}\footnote{BAT has earlier coverage of AT2018cow from its non-pointed survey data}, the X-Ray Telescope \citep[XRT]{2005SSRv..120..165B} and the UltraViolet \& Optical Telescope \citep[UVOT]{2005SSRv..120...95R}, on MJD~58288.44 which was 3.0\,d after the first detection, and {\revv 
continued with an intensive observing schedule over the following 2 months. }
Unless said otherwise, the \Swift data were reduced using {\tt HEAsoft}-6.22 (XRT), 6.24 (UVOT) and the latest {\em Swift}\,{\tt CALDB} or, for the UVOT grism data, with the {\tt uvotpy} calibration and software \citep{2015MNRAS.449.2514K}.
The XRT spectra were obtained using the online XRT product
generator at the UK Swift Science Data Centre 
 \cite[][]{2009MNRAS.397.1177Evans}. 
 We used the Galactic absorption N$_H$ from \citet[][]{2013MNRAS.431..394Willingdale}.

\begin{figure*}
\includegraphics[width=14.0cm]{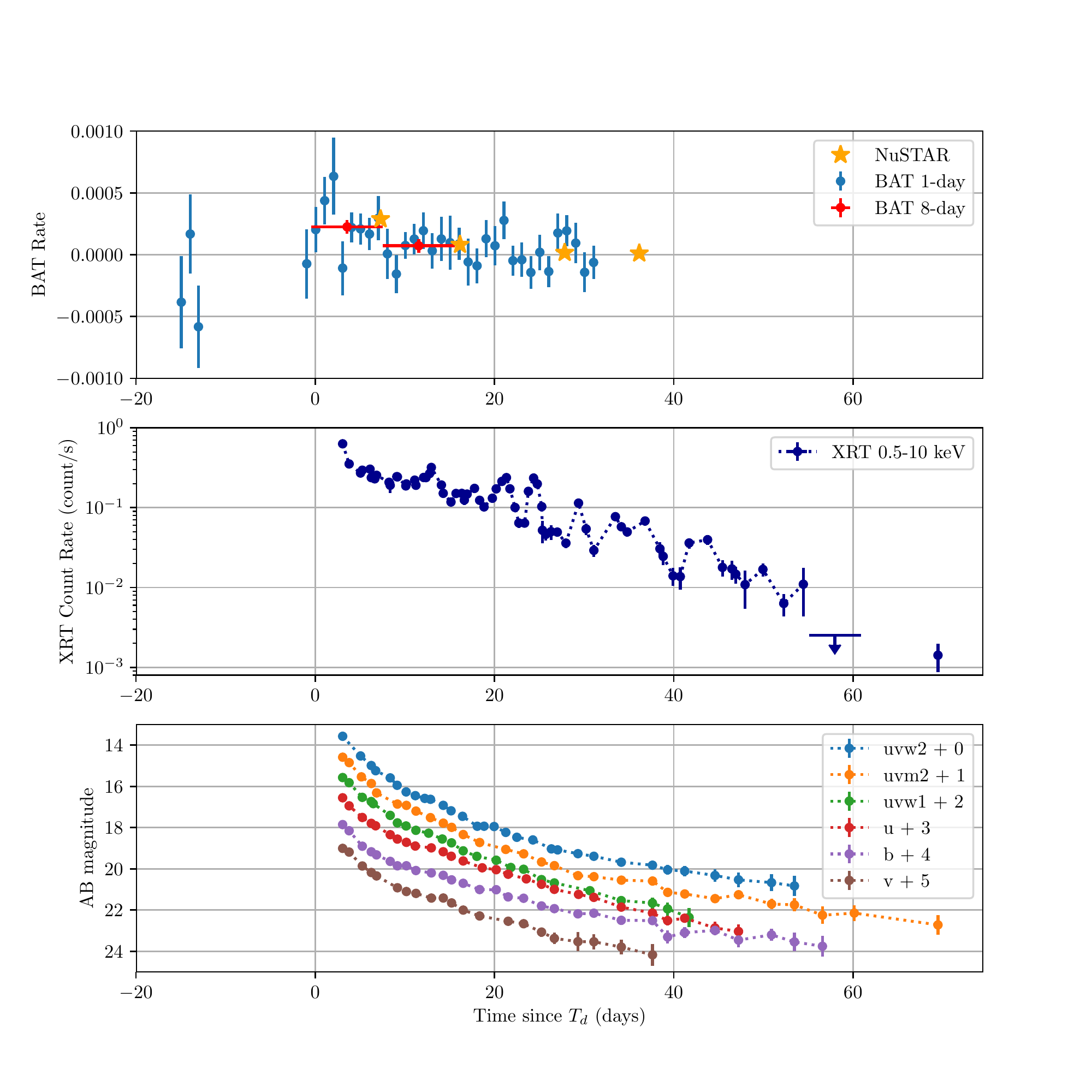}
\caption{BAT-XRT-UVOT light curve. 
The UVOT magnitudes are given in six filters $uvw2, uvm2, uvw1, u, b$, and $v$ starting at \Td~ and have been corrected for the galaxy background and have been binned to increase S/N.
The BAT survey data panel includes the NuSTAR data projected into the BAT band, as well as the BAT survey quality processed data for 8-day periods. 
During the first 8-day period significant detections occur, thereafter the BAT count rate is consistent with no detection.
The flaring seen in the XRT possibly lines up with an increase in the BAT flux prior to day \Td+8. 
}
\label{fig:BAT_survey}
\label{fig:UVOT_lc_Tdisc}
\label{fig:XRT_lc} 
\end{figure*}

\subsection{The UV-optical light curve, and SED}
\label{sec:uvot}

The UVOT images were inspected for anomalies, like drift during the exposure. 
Photometry was obtained using the standard {\tt HEAsoft}-6.24 tools followed by a check that the source did not fall on one of the patches of reduced sensitivity; several observations had to be discarded. 
A 3\arcsec~ aperture was used throughout; the standard aperture correction has been used as described in \citet{Poole2008}; and the filter effective areas and zeropoints were from \citet{Breeveld2011}. 
For the fitting of the photometry in {\tt Xspec} the UVOT filter response curves were used so that the fits take the interplay between filter transmission and spectrum into account.

The galaxy emission in the aperture of 3\arcsec~ radius was determined from a 
{\revv UVOT observation  at day 120, in order to correct the UVOT photometry for the host contribution. The galaxy background was measured in all 6 UV/optical filters, but the transient was still judged to dominate the UVOT emission in the UV.
The UV background from the galaxy was estimated by determining that from } 
the GALEX FUV and NUV filters using the {\tt gPhoton} database \citep[][]{2016ApJ...833..292M_gphoton}.
There is a blue source near the transient which falls within the 3\arcsec\, aperture. 
The GALEX PSF is larger than UVOT, so the bright source and contribution from the bulge of the galaxy will lead to an overestimate of the flux in the 3\arcsec\,  radius aperture used. 
An SED was built for the galaxy emission component and folded through the UVOT effective area curves to derive the following galaxy flux and magnitude within the aperture: 
{\revv $uvw2$=20.46, $uvm2$=20.37, $uvw1$=19.70 (AB). 
The host galaxy values from UVOT data taken at day 120 were   $u$=19.14, $b$=18.50, and  $v$=17.92 (AB) to an accuracy of 0.08 mag.} 
The galaxy emission becomes important first in the UVOT $v$ band around day 12, and later in the bluer bands.
The UVOT photometry corrected for the galaxy background as described above can be found in Table~\ref{tab:uvot_photometry}.
The UVOT light curves 
show a chromatic decline, where the $uvw2$, and $uvm2$ fall off slower than the optical $u$, $b$ and $v$ bands, see Fig.~\ref{fig:UVOT_lc_Tdisc}.

\begin{figure}
\includegraphics[width=\columnwidth]
{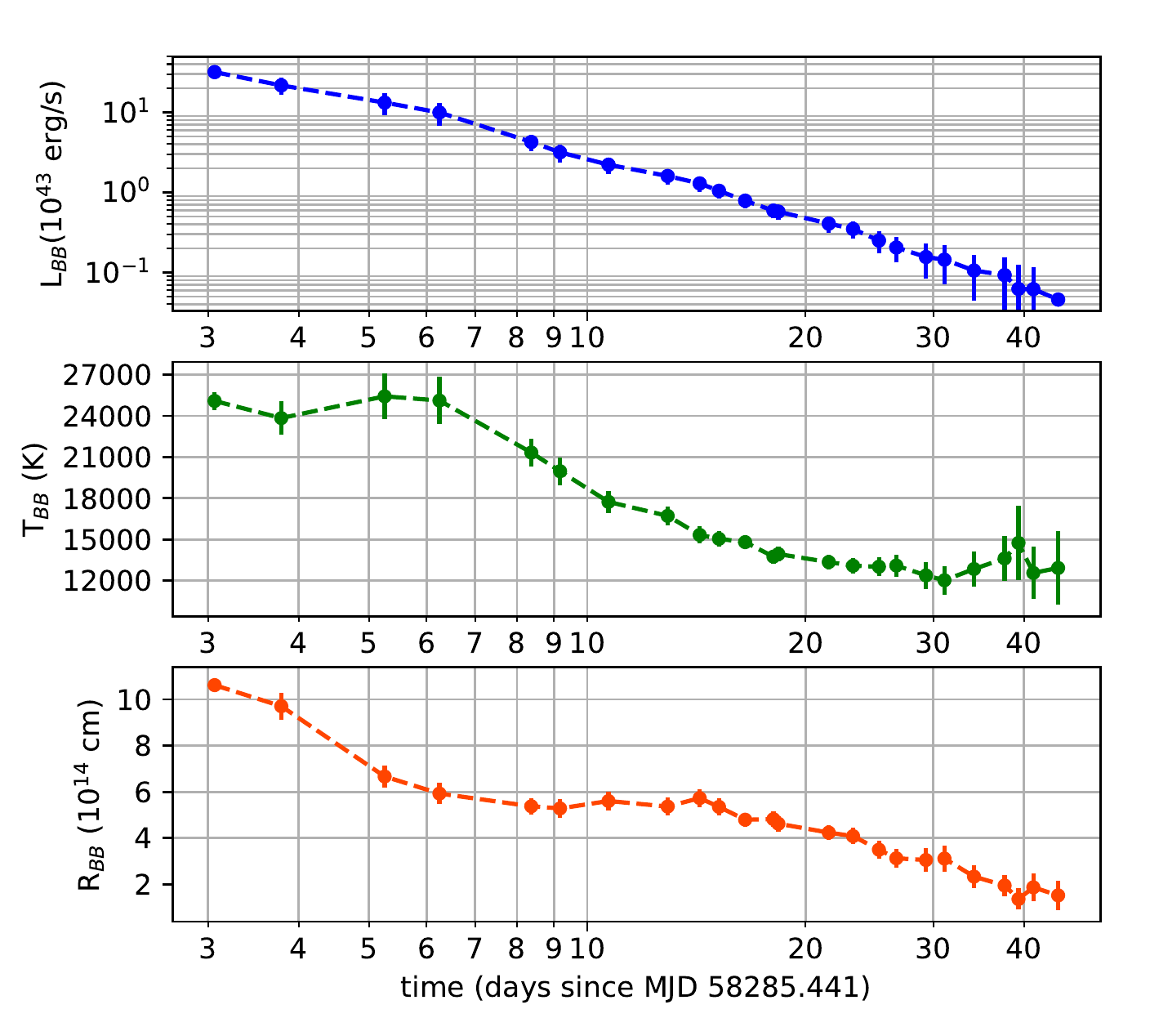}
\caption{Black-body model fit to the UVOT data in a log-log plot. Note that the luminosity decays approximately as a power law with different slopes before and after day \Td+6.5. After day \Td+44 the galaxy background emission in $v$ and $b$ do no longer allow a good fit to be made.} 
\label{fig:BBfit}
\end{figure}

The first report that the optical-IR spectra resembled a blackbody was by \citet{2018ATel11729....1C}  who reported a temperature of $9200\pm 600$\,K on day \Td+1.7\,d.
A more detailed fit was made by \citet{2018arXiv180705965P} who discuss the UV-optical/infrared data from the first 17\,d after discovery. 
They also showed that the data can be fit well with a blackbody spectrum, with luminosity changing over an order of magnitude, blackbody  temperature changing from 28,000\,K to 14,000\,K and 
a nearly constant radius of the photosphere at $5\times 10^{14}\,{\rm cm}$. 
\citet[][]{2018arXiv180800969Perley} subsequently modeled ground based photometry and spectra, and reported a slow but  steady decline of the photospheric radius, but they also needed a power law component to fit excess emission in the infrared. 
{\revv The IR excess emission could be synchrotron emission from non-thermal energetic electrons present in an optically thin coronal  atmosphere above a dense photosphere. 
Perley at al. (2018a) noted that 
 such IR synchrotron emitting electrons 
 could produce radio synchrotron emission 
 at a level consistent with the observation.  
Analyses by Ho et al. (2018) showed further support 
  to the scenario that the IR excess emission 
  and the radio emission are of the same origin. }
Using {\tt Xspec} we fit a blackbody model to our corrected UVOT photometry. 
The results are given in Table~\ref{tab:uvot_bbmodel} and Fig.~\ref{fig:BBfit}.
Our analysis is confined to the well-calibrated UVOT data, with bad data removed, taking account of the filter throughput with wavelength, and correcting for the measured galaxy background. 
However, the $v$ magnitudes are generally too bright due to the extra red power law emission component, and lead to a poor reduced $\chi^2$. 
We see a varied evolution during the first 13\,d, see  Fig.\,\ref{fig:BBfit}, followed by a steady decline in radius.

\begin{table}
\caption{{\revv Results of the black-body model fits} to the UVOT photometry (1700 - 6800\,\AA)}
\label{tab:uvot_bbmodel}
\begin{tabular}{ccccrccc}
\hline
Time      & T$_{\rm BB}$     & $R_{\rm BB}$   & $L_{\rm BB}$ & $\chi^2$/d.o.f.\\
(days) & (10$^3$\,K) & ($10^{14}$\,cm)  & (10$^{43}\,{\rm erg\,s}^{-1}$)   \\
\hline
 3.06 &$ 25086 \pm  650 $&$10.62 \pm 0.27 $&$ 31.80 \pm 3.67 $& 23.33/4.0 \\
 3.78 &$ 23840 \pm 1227 $&$ 9.71 \pm 0.58 $&$ 21.69 \pm 5.16 $& 17.48/4.0 \\
 5.25 &$ 25424 \pm 1657 $&$ 6.67 \pm 0.48 $&$ 13.22 \pm 3.95 $& 25.54/4.0 \\
 6.25 &$ 25117 \pm 1712 $&$ 5.93 \pm 0.46 $&$  9.96 \pm 3.11 $& 21.29/4.0 \\
 8.37 &$ 21328 \pm 1013 $&$ 5.37 \pm 0.36 $&$  4.26 \pm 0.99 $& 17.44/4.0 \\
 9.17 &$ 19973 \pm 1004 $&$ 5.28 \pm 0.39 $&$  3.16 \pm 0.79 $& 6.79/4.0 \\
10.70 &$ 17738 \pm  804 $&$ 5.60 \pm 0.40 $&$  2.21 \pm 0.51 $& 12.97/4.0 \\
12.91 &$ 16719 \pm  698 $&$ 5.36 \pm 0.38 $&$  1.60 \pm 0.35 $& 10.37/4.0 \\
14.29 &$ 15334 \pm  620 $&$ 5.73 \pm 0.40 $&$  1.29 \pm 0.28 $& 8.41/4.0 \\
15.19 &$ 15050 \pm  537 $&$ 5.35 \pm 0.36 $&$  1.05 \pm 0.21 $& 7.06/4.0 \\
16.51 &$ 14809 \pm  507 $&$ 4.79 \pm 0.31 $&$  0.79 \pm 0.15 $& 15.1/4.0 \\
18.06 &$ 13751 \pm  496 $&$ 4.82 \pm 0.34 $&$  0.59 \pm 0.12 $& 17.99/4.0 \\
18.35 &$ 13939 \pm  542 $&$ 4.62 \pm 0.34 $&$  0.57 \pm 0.12 $& 15.51/4.0 \\
21.53 &$ 13355 \pm  533 $&$ 4.24 \pm 0.33 $&$  0.41 \pm 0.09 $& 13.39/4.0 \\
23.25 &$ 13090 \pm  571 $&$ 4.08 \pm 0.35 $&$  0.35 \pm 0.09 $& 5.52/4.0 \\
25.25 &$ 13018 \pm  696 $&$ 3.50 \pm 0.38 $&$  0.25 \pm 0.08 $& 12.48/4.0 \\
26.68 &$ 13101 \pm  805 $&$ 3.12 \pm 0.39 $&$  0.21 \pm 0.07 $& 8.23/4.0 \\
29.31 &$ 12383 \pm  996 $&$ 3.05 \pm 0.51 $&$  0.16 \pm 0.07 $& 6.35/4.0 \\
31.09 &$ 12031 \pm 1042 $&$ 3.11 \pm 0.57 $&$  0.15 \pm 0.07 $& 5.48/4.0 \\
34.15 &$ 12851 \pm 1296 $&$ 2.33 \pm 0.48 $&$  0.11 \pm 0.06 $& 5.99/4.0 \\
37.63 &$ 13618 \pm 1641 $&$ 1.95 \pm 0.46 $&$  0.093 \pm 0.063 $& 11.08/4.0 \\
39.32 &$ 14753 \pm 2700 $&$ 1.36 \pm 0.46 $&$  0.062 \pm 0.062 $& 1.5/4.0 \\
41.22 &$ 12567 \pm 1902 $&$ 1.87 \pm 0.59 $&$  0.062 \pm 0.054 $& 4.29/4.0 \\
44.60 &$ 12928 \pm 2685 $&$ 1.52 \pm 0.64 $&$  0.046 \pm 0.054 $& 12.21/4.0 \\
\hline
\end{tabular}
\end{table}

\subsection{The UVOT Spectra}

\begin{figure}
\includegraphics[width=\columnwidth]{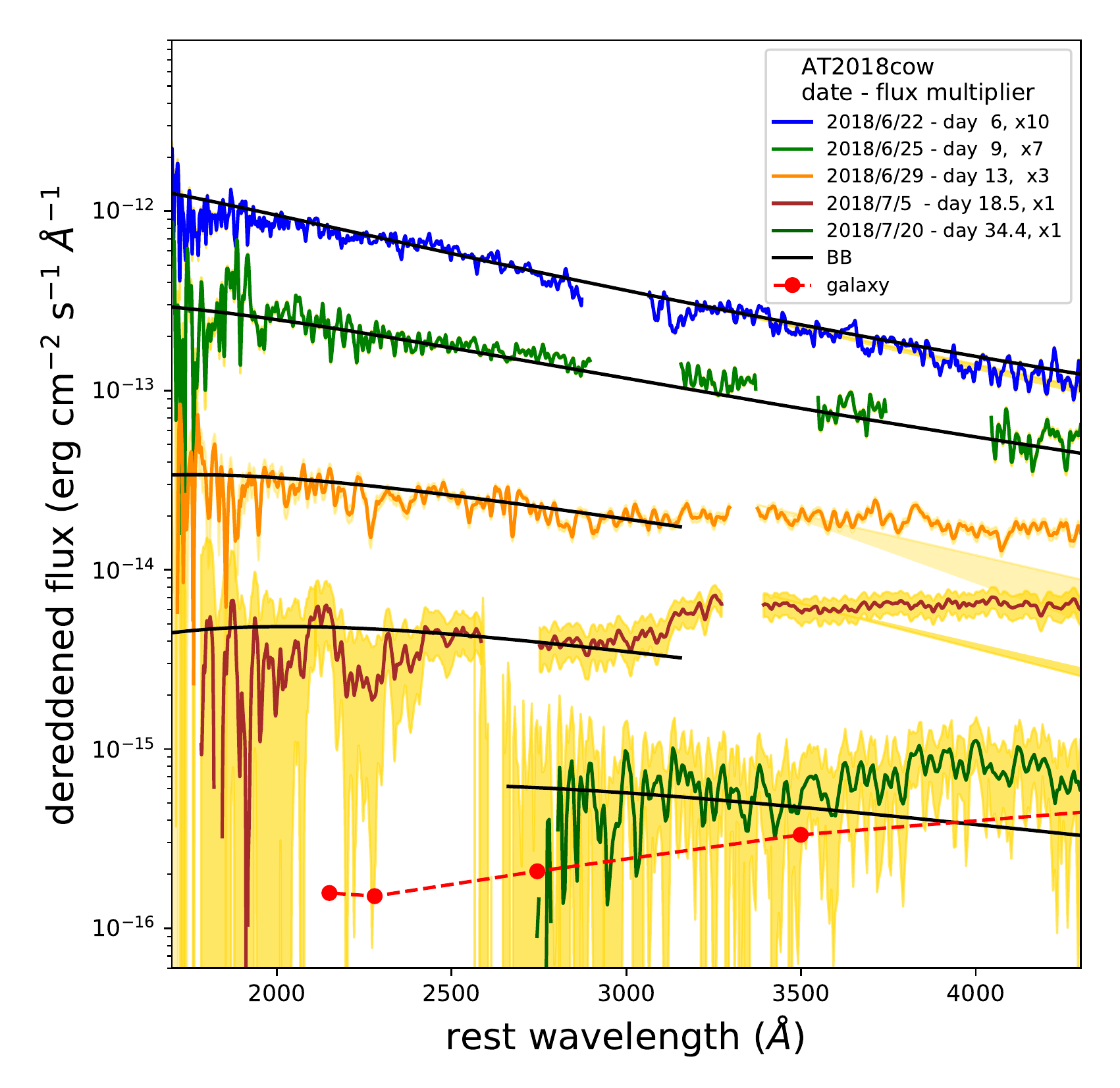}
\caption{The dereddened UVOT grism spectra, using $E(B-V) = 0.077$ and the  
\protect\citet{1989ApJ...345..245C} law with $R_V = 3.1$.
The 1$\sigma$ errors are indicated with shading. 
The features below 2200\,\AA\  are due to noise. 
Black-body fits are included, as well as the galaxy brightness in the 3\arcsec~  radius aperture used for photometry. 
Excess flux on day \Td +13 and 18.5 longer than 3000\,\AA\ is due to  order overlap.}
\label{fig:uvot_spectra}
\end{figure}

Daily exposures in the UVOT grisms were obtained from \Td+5\,d onward, first in the UV\,grism ($170-430$\,nm) until day \Td+23.4\,d; on day \Td+24, 25, 30 and 34\,d  we obtained exposures in the more sensitive V\,grism ($270-620$\,nm).  
The UVOT grism images were closely examined for contamination by background sources using the summed UVOT UV filter images as well as by comparison of the position of the spectrum to zeroth orders from sources in images from the Digital Sky Survey. 
The affected parts of the spectrum were removed from consideration. 
To improve S/N, spectra were extracted with a narrow slit measuring 1.3 times the FWHM of a fitted Gaussian across the dispersion direction, and the extracted spectra taken close in time were averaged together. 
A standard correction to the flux from the narrow slit was made to scale it to the calibrated response \cite[see][]{2015MNRAS.449.2514K} and a correction was made to account for the coincidence loss in the detector. 
The resulting spectra (see Fig.\,\ref{fig:uvot_spectra}) show little evidence for emission lines like in, for example, novae, nor the characteristic UV  absorption features due to blended lines of singly ionised metals as seen in SNe. 

{\revv Our first summed spectra from around day 6 are well-exposed, yet relatively featureless,  just like the optical spectra in \cite{2018arXiv180705965P}. }
The dereddening straightened the bump in the observed spectra around 2175\,\AA~ which means that there is no evidence of dust intrinsic to the environment of the transient.  
The June 25, day \Td +9, spectrum shows features near 1910\AA, which are probably due to a noise problem, since this feature would likely have been seen as  second order emission.
Therefore the features are probably unrelated to the broad 4850\,\AA\  absorption (or broad emission around 5000\,\AA)  feature which is seen to emerge in the ground-based spectra from \citet[][]{2018arXiv180800969Perley} on day $9-13$.
Our UV-grism spectra from June 27.5 to July 1.6 were combined, to give a spectrum for day 13 (June $29.5\pm 2$~d). In this spectrum a weak emission feature near the He\,II 2511\,\AA\  line is seen, as well as a broad feature near 2710\,\AA\  which could be due to He\,I or He\,II. 
At wavelengths longer than 3000\,\AA\ second order overlap contamination is present. 
We should note that the second order lines of N\,III] and C\,III] would be seen if there were any, and their absence in second order confirms that there is no line in the first order.
The UV spectrum from day 18.5 is contaminated by second order emission overlap for wavelengths longer than about 3300\,\AA, while below 2400\,\AA\  noise starts to 
dominate {\revv \citep{2015MNRAS.449.2514K}. 
The dip near 2200\,\AA~ is likely due to noise. }

At day \Td+34 we got {V-grism} exposures which were of low signal to noise. 
{\revv Usually, the UVOT spectra are extracted from each individual exposure and then the wavelength reference is corrected to match the spectra before summing. 
However, the latter is not possible if the spectrum in each exposure is too weak.
An alternative is possible for exposures taken with the same spacecraft roll angle.}
To get a better S/N for the day \Td+34 spectrum, we cross-correlated the images and then summed the grism images of the spectrum, followed by a standard extraction, again using the narrow extraction slit {\revv \citep{kuin2014}}. 
The spectrum of day 34, which covers the range of 2820-5600\,\AA, shows undulations which resemble those shown for the longer wavelengths in the spectra from \citet[][]{2018arXiv180800969Perley}. 
Below 3800\,\AA, which was not covered in the ground-based spectra, we see no evidence for strong emission lines from other elements, i.e., the  Mg\,II~2800\,\AA\  emission line 
{\revv which is often found in late type stellar spectra} is not seen; nor is there any sign of the O\,III~3134\,\AA\ line which is pumped by He\,II~304\,\AA~ 
{\revv nor of He\,II~3204}.
Using the BB-fits to our photometry, we  plot those over the UVOT spectra in Fig.~\ref{fig:uvot_spectra}.

Whereas the UVOT spectra become quite noisy after day \Td+20 and do not show any lines above the noise, it is of interest to mention that the {\revv He\,I} spectral lines that emerge after $\sim$~day~22 in the ground-based spectra from \citet[][{\revv figure 4}]{2018arXiv180800969Perley} show large asymmetries.  
{\revv  The blue wing is seen to be largely missing in the stronger unblended lines. 
After the lines become visible above the continuum, the line emission at first peaks at a 3000 km s$^{-1}$ redshift, while moving to lower redshifts until the peak is at the rest wavelength at day 34}.
We think those asymmetric line profiles are important for understanding the transient.

\begin{figure}
\vspace{-29mm)}
\includegraphics[width=\columnwidth]{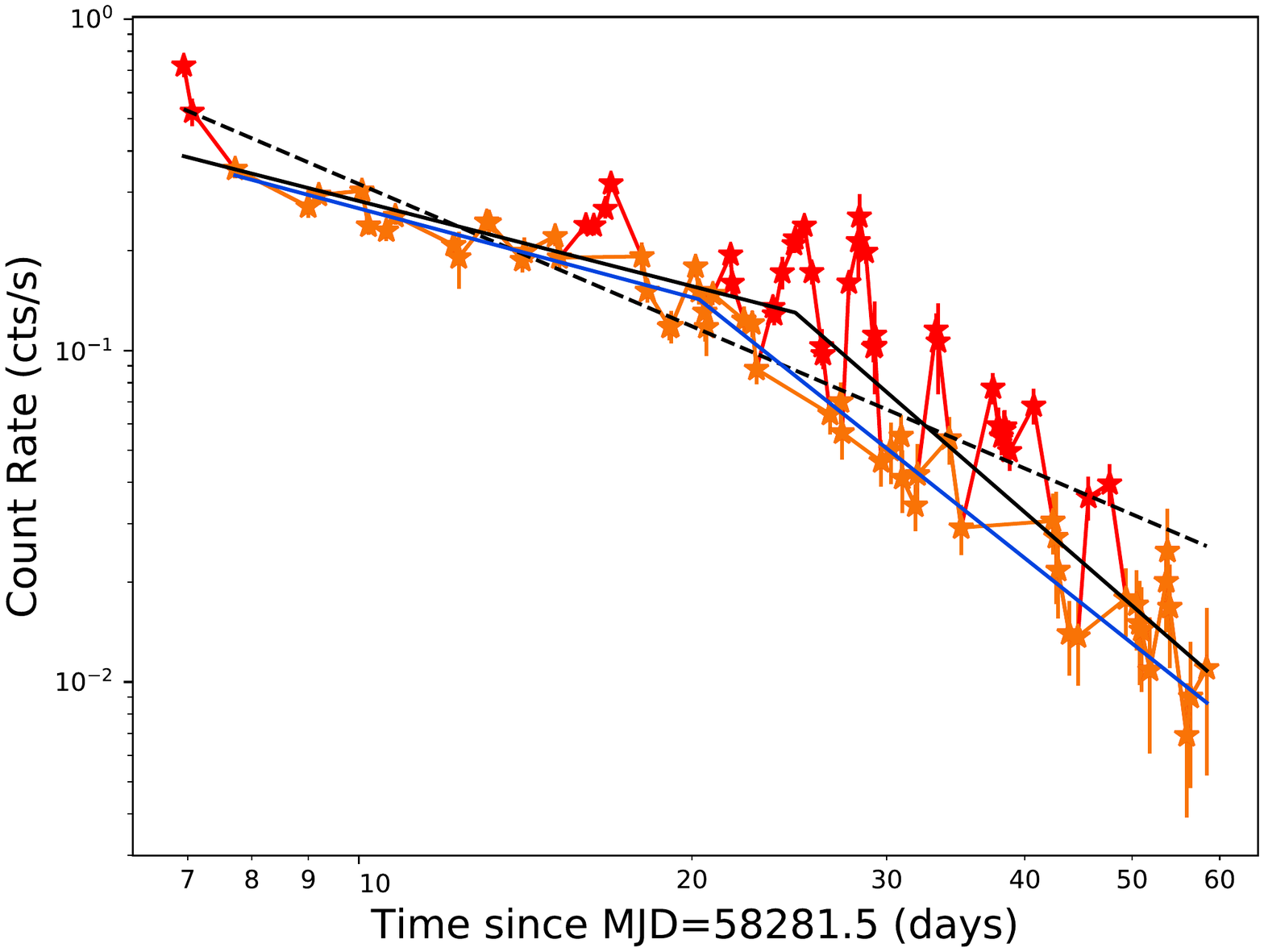}
\vspace{-30mm}
\caption{
A log-log plot of the XRT light curve indicates that the evolution of the X-ray emission follows a broken power-law with a break at day \Td+21.
Three fits are shown: fitting all data with a single PL (black dashed line), or with a broken PL (black line) as well as a fit to the data excluding obvious flaring points (blue line).
The flaring points which are shown here in red.}
\label{fig:XRT_lc_loglog}
\end{figure}

\subsection{XRT analysis}

The X-ray light curve shows a regular pattern of brightenings which have already been remarked upon by \cite{2018arXiv180706369R}. 
These rebrightenings follow a trend, either as flares above a certain base-level, or as the main constituent. 
A simple PL fits to  all data has $\alpha = -1.43\pm 0.08$ for a $\chi^2$/d.o.f. = 1316.1/87. 
Alternatively, fitting all the data including the flares with a broken PL has a slope $\alpha_1$ = $0.85\pm 0.11$, with a break in its 
slope at day $t_{\rm break}$ = \Td$+24.8 \pm 1.6$, followed by a 
steeper decay with a slope $\alpha_2$ = $2.90\pm0.35$, and $\chi^2$/d.o.f. = 803.7/85.
{\revv We also made a fit trying to find a trend underlying the flares.}
A base level can be defined by masking the points in a flare, fitting a broken power-law (PL) trend, and iteratively removing points that are too far from the trend by eye 
(see Fig.\,\ref{fig:XRT_lc_loglog}).
The best fit found is a broken power law with a slope $\alpha_1$ = $0.89^{+0.22}_{-0.23}$, which has a break in its slope at day 
$t_{\rm break}$ = \Td+20.3$^{+2.7}_{-3.8}$, followed by a steeper decay with a slope $\alpha_2$ = $2.65^{+0.57}_{-0.43}$, and 
$\chi^2$/d.o.f. = 89.3/50.
In Fig.\ref{fig:XRT_lc_loglog} one can clearly see that the single power law does not fit as well at the beginning and end. 
The same trend is thus present regardless of the removal of the flares. 
Although the flares seem to be on top of a smoothly varying component, this cannot be determined for sure since the whole overall emission is decreasing over time, and it could just as well be continuous flaring which shows evolution.  
We will investigate the temporal behaviour of the flares further in section \ref{sec:timing}.

\subsection{BAT analysis}
\label{sec:BAT}

{\revv BAT is a coded aperture imaging instrument (Barthelmy et al. 2005). A sky image can be constructed by deconcolving the detector plane image with the BAT mask aperture map \citep{Markwardt07}. We performed a special analysis of the BAT data.}
This analysis utilizes the BAT survey data from  June 1st to July 18th, 2018 (i.e., the available HEASARC data at the time of the analysis). 
Even when the BAT has not been triggered by a GRB, it collects continuous survey data with time bins of $\sim 300$\,s \citep[see detailed descriptions in][]{Markwardt07}. 

{\revv 
 The signal to noise ratios reported here are calculated using the source count and background variation estimated from sky images with different exposure time \cite[for details see ][]{Tueller10}. Note that due to the nature of the deconvolution technique, the resulting noise (background variation) is Gaussian instead of Poissonian. These sky images are mosaic images created by adding up all the snapshot observations within the desired durations (i.e., two 8-day intervals and one 17-day period). The mosaic technique adopted here is the same one that is used to create the BAT survey catalogs \cite[][]{Tueller10,  Baumgartner13, Oh18}. 
}
This analysis pipeline carefully takes care of many instrumental effects, such as potential contamination from bright sources, systematic noise introduced by differences between each detector (so-called ``pattern noise''), and corrections for sources with a different partial coding fraction when creating a mosaic image from individual snapshot observations.

The analysis produces results in the following eight energy bands:  $14-20$\,keV, $20-24$\,keV, $24-35$\,keV, $35-50$\,keV, $50-75$\,keV, $75-100$\,keV, $100-150$\,keV, and $150-195$\,keV.

Figure \ref{fig:BAT_survey} shows the daily BAT mask-weighted light curve in $14-195$\,keV.
Note that we exclude data collected from June 3rd to June 13th, 2018, 
during which the BAT underwent maintenance and recovery activities and the calibration of survey data is
uncertain.

A spectrum was created for the 17-d period starting June 16 (day 0) to July 2, 2018, as well as for two 8-day periods of June 16 $-$ June 23, 2018, and June 23 - July 1, 2018. 
For the 17-d period the S/N was 3.3, and for the two 8-d periods, 3.95 and 1.31 respectively. 
For new sources the BAT detection limit is a higher level of S/N of 5, so these $3-4\;\!\sigma$ detections are marginal detections that do not stand on their own. 

\subsection{The high energy spectra}

The {\em Swift} XRT data up to day 27 have been discussed in \cite{2018arXiv180706369R} who fit an absorbed power law to the data. 
Their spectral fits did not show any evidence for spectral evolution in the $0.3-10$\,keV band, and no evidence for spectral evolution during the flares. 
The latter is also consistent with no changes being seen in the hardness ratio. 
Their estimate for the peak X-ray luminosity is $10^{43}\,{\rm erg\,s}^{-1}$.

We addressed the UVOT data in detail in section~\ref{sec:uvot} but here we want to address {\revv the question of whether the origin of the} X-ray emission is related to the UV emission.
The emission in the UV-optical is well fitted by a hot thermal blackbody (BB)
{\revv which we interpreted as }
optically thick emission from an ionised He sphere surrounding the source.
The optical luminosity of $\sim 2 \times 10^{44}$\,erg s$^{-1}$  \citep[see Table\,\ref{tab:uvot_bbmodel} and][]{2018arXiv180705965P} is larger than the X-ray luminosity \citep{2018arXiv180706369R}, and Compton scattering on the electrons in the atmosphere may produce an X-ray spectrum, so we investigate if that would be large enough to explain the observed X-ray luminosity. 
We used the {\tt XSPEC} tool to model the UVOT photometry together with the XRT data.  
We used the data on day \Td+21 (using the data from UVOT: MJD~58306.8; 
 XRT MJD~$58306.2\pm 1.0$) to determine if the optically thick Compton scattering of the BB spectrum was consistent with the observed X-rays.
Since a simple BB gives a reasonable fit, the optical depth of the scattering atmosphere was set to one. 
Fitting the combined UV-optical and X-ray spectral data with an optically thick ($\tau \approx 1$) Compton spectrum 
{\tt (compbb*zphabs*redden*phabs)} fails to get a reasonable fit for the X-ray data which are underestimated: its 
 $\chi^2 = 145.69\,(64\,{\rm d.o.f.})$.
 This suggests that the X-rays are not due to the same source as the blackbody emission. 

A much better fit is obtained using a model of optically thick Compton scattered BB plus a power law (PL),
{\tt (compbb + powerlaw)*redden*phabs*zphabs} with  
 $\chi^2 = 54.0\,(61\,{\rm d.o.f.})$ and which  gives results very close to a BB+PL model with similar BB temperatures as in \cite{2018arXiv180705965P} for the low-energy part of the spectrum. 
The photon index $\beta$ of the day \Td+21 unabsorbed X-ray spectrum  is  $\beta = 1.57 \pm 0.07$\footnote{We parametrise the photon index as $n_{\nu} \sim t^{-\alpha}\nu^{-\beta}$ }.


We fitted a broken power law spectrum to the combined BAT and XRT spectral data in order to determine the high-energy losses. 
We used the summed BAT data for the first 8-day interval of \Td+0 to 8\,d, and for XRT day \Td +3 to 8\,d. 
Assuming that this represents a detection, we obtain for the broken power law fit a photon index $\beta_1 = 1.792 \pm 0.065$, a break at $6.57\pm 1.64$\,keV, 
and thereafter $\beta_2 = 0.65\pm 0.13$, 
  with a fixed $N_{\rm H,galactic} = 6.57\times 10^{20}\,{\rm cm}^{-2}$, and 
$N_{\rm H,intrinsic} = (1.9 \pm 1.5)\times 10^{20} {\rm  cm}^{-2}$. 
The model predicts a luminosity ratio between the BAT and XRT bands 
  of $L_{\rm x}(10-200\,{\rm keV})/L_{\rm x}(0.3-10\,{\rm keV}) = 29.8$, 
    while the goodness of fit 
$\chi^2 = 82.4\,(90 {\rm d.o.f.})$.

\begin{figure}
\includegraphics[width=\columnwidth]{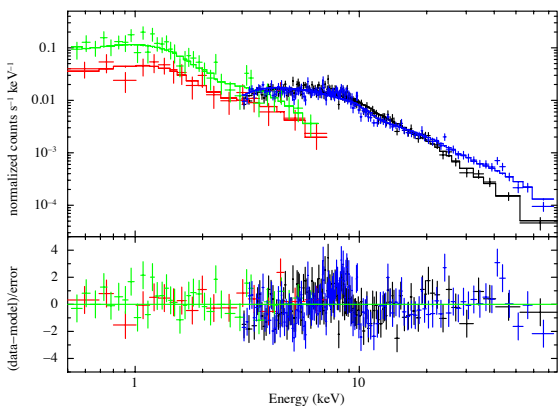}
\caption{The combined XRT and NuSTAR spectra fit. The first NuSTAR observation (blue) and the 2nd NuSTAR observation (black) are shown with their XRT counterparts (green and red respectively).}
\label{fig:XRT_NuSTAR_spectra}
\end{figure}

In addition to {\em Swift} observations, NuSTAR \citep{NuSTAR} observed AT2018cow on four occassions during the period reported upon here.  NuSTAR observes in an energy range of 3-79~keV, overlapping the energy range of both XRT and BAT. As NuSTAR has a greater sensitivity than BAT, it allows us to validate the quality of our XRT + BAT spectral fit. 
We analysed the NuSTAR data utilizing the standard extraction methods, utilizing the HEAsoft {\texttt{nupipeline}} and {\texttt{nuproducts}} tools, and in cases where XRT and NuSTAR observed simultaneously, we include both data in the fit. 

The first NuSTAR observation occurred on MJD 58292.7, $\sim7.3$ days after discovery. 
By simultaneously fitting XRT and NuSTAR data we find that, similar to the previously reported BAT + XRT fit, it requires a broken power-law model to fit the data. 
This fit gives $\beta_1 = 1.67 \pm 0.04$, with a spectral break at $12.7 \pm 0.9$~keV, followed by $\beta_2 = 0.50 \pm 0.10$. 
This model adequately ($\chi^2$/d.o.f. = 572.6/511) describes the data, see Fig.\,\ref{fig:XRT_NuSTAR_spectra}. We note that although the spectral indices are similar to the BAT + XRT fit, the energy at which the spectral break occurs is higher, and importantly, outside of the XRT 0.3 - 10 keV energy range. It is not clear if this difference is instrumental or due to the different time periods over which the spectra were collected. Utilizing this fit we derive a luminosity ratio between the BAT and XRT bands of $L_x$(10-200 keV)/$L_x$(0.3-10 keV) = 18.6.

Further NuSTAR spectra taken at \Td +16.2, 27.8 and 36.15 show a marked change in the spectral shape, as the hard component disappears and the combined XRT and NuSTAR data for each observation can be well fit by a single power-law model from 0.5 to 79 keV. Based on the spectral softening seen in NuSTAR, it is clear that AT2018cow would not have been detected by BAT after this hard component turned off. However the relatively low statistics of the BAT light-curve means that it is not possible to estimate when this hard component turned off, although we note that after \Td +8 days there are no statistically significant detections of AT2018cow by BAT either in 1-day or 8-day integrated data. 
{\revv These results are consistent with the report of a ``hard X-ray bump" at day 7.7 which disappears by day 17 by \citet{2018arXiv181010720M_Raf}. }

\subsection{Search for characteristic or periodic time scales} 
\label{sec:timing}

In the XRT photon-counting (PC) data there are some flares visible. 
To investigate whether these were periodic, semi-periodic, or burst-like, we 
{\revv conducted a structure function analysis, following the prescription described in \citet[][]{2012MNRAS.422.1625S}. The analysis reveals a weak, not very significant, characteristic time scale around 4 hours for count rates binned on 100~s. }

A Lomb-Scargle period search was repeated in several ways,  initially by using data binned per orbit and binned every 100s. We determined the probability that the power of the periodogram was obtained by chance. 
We simulated the  light curve using the same {\revv observation} times as the original data, but resampling the count rate with replacement in a Monte Carlo simulation. After $10^5$ iterations the values at $3\sigma$ and $5\sigma$ were extracted from the resulting power distribution at each frequency.

Periods of about 3.7\,d and 90\,min are found in the periodograms, above the $5\sigma$ level for the data binned per orbit and every 100s. The 90\,min period is due to the {\em Swift} orbit. In order to determine if the long period is robust, we repeated the analysis using detrended data, i.e., by excluding the long term trend using a broken PL fit. 
The analysis was repeated and found consistent results.
However, when we split the detrended XRT data into three equal time segments and repeated the Monte Carlo Lomb-Scargle analysis, we find that while a consistent period is found in the latter two thirds of the data, this period is not present in the first third of the data, suggesting the period is quasi-periodic/temporary and not an inherent property of the system.
A Monte-Carlo Lomb-Scargle period search in the $uvw2$ band shows no evidence for periodicity except at the orbital period of the 
 {\em Swift} satellite. 

\begin{figure}
\includegraphics[width=\columnwidth,angle=0]{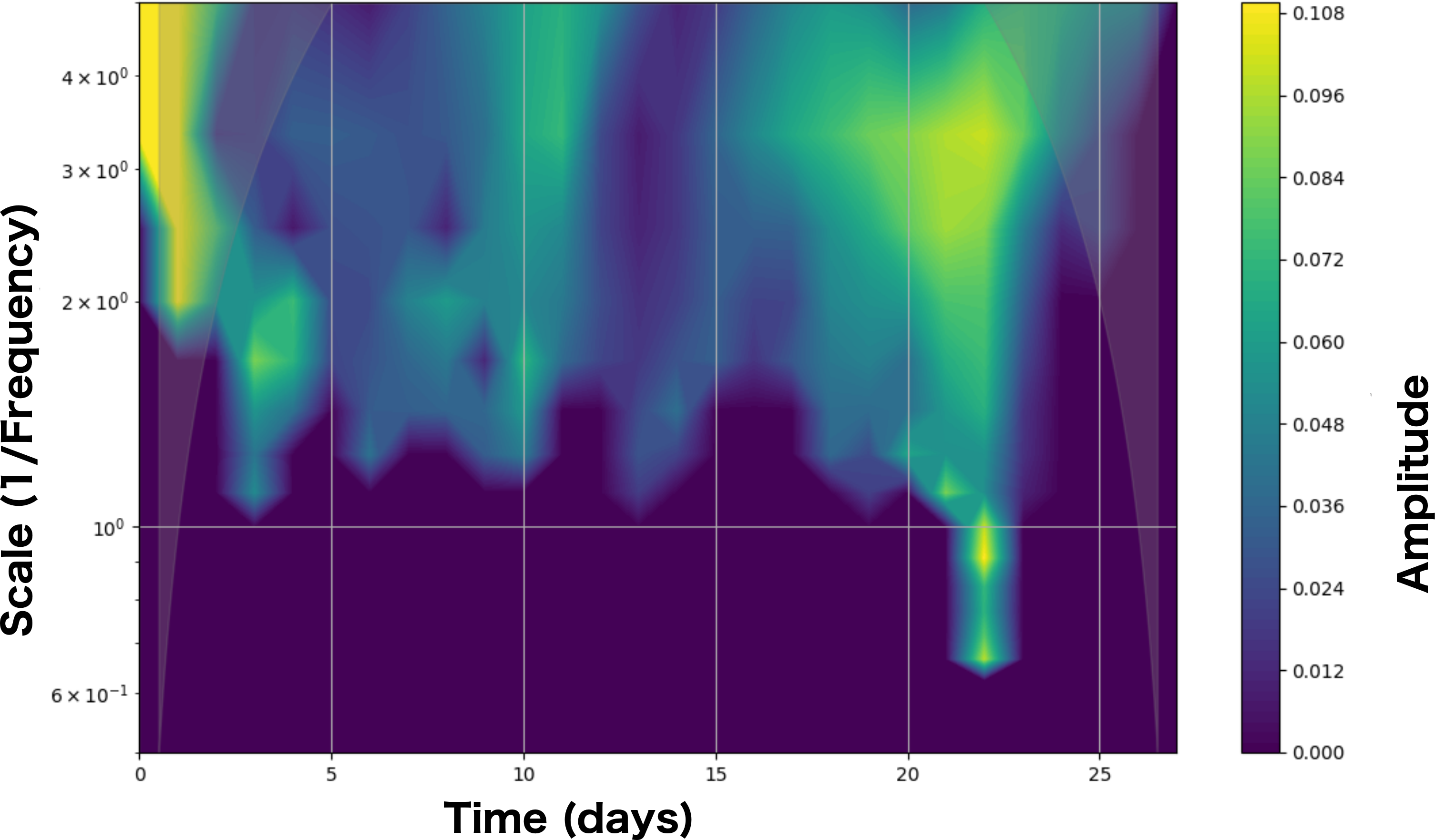}
\caption{Wavelet analysis of the XRT PC data shows no period but bursty behaviour. The Y-axis shows the time-scales (in days) searched for.}
\label{fig:wavelet}
\end{figure}

Finally, we performed a {\em wavelet} analysis \citep{1996AJ....112.1709F} of the XRT data. 
We found no periodic signal, see Fig.\ref{fig:wavelet}; instead we see that at certain times there is a burst of activity. 
To investigate whether the variability is wave- or burst-like we also calculated Pearson's moment coefficient of skewness of the amplitude. 
We disregarded the first ten data points, which cause a large skew{, \revv much larger than the rest of the time series presents because of the initial large drop in brightness}. 
The changes in count rate over the mean trend are  significantly skewed with a coefficient of 0.67,   which shows that the brightenings are burst-like, not periodic.

\section{A model of tidal disruption}

\subsection{AT2018cow compared to other transients}

\begin{figure}
\includegraphics[width=\columnwidth]{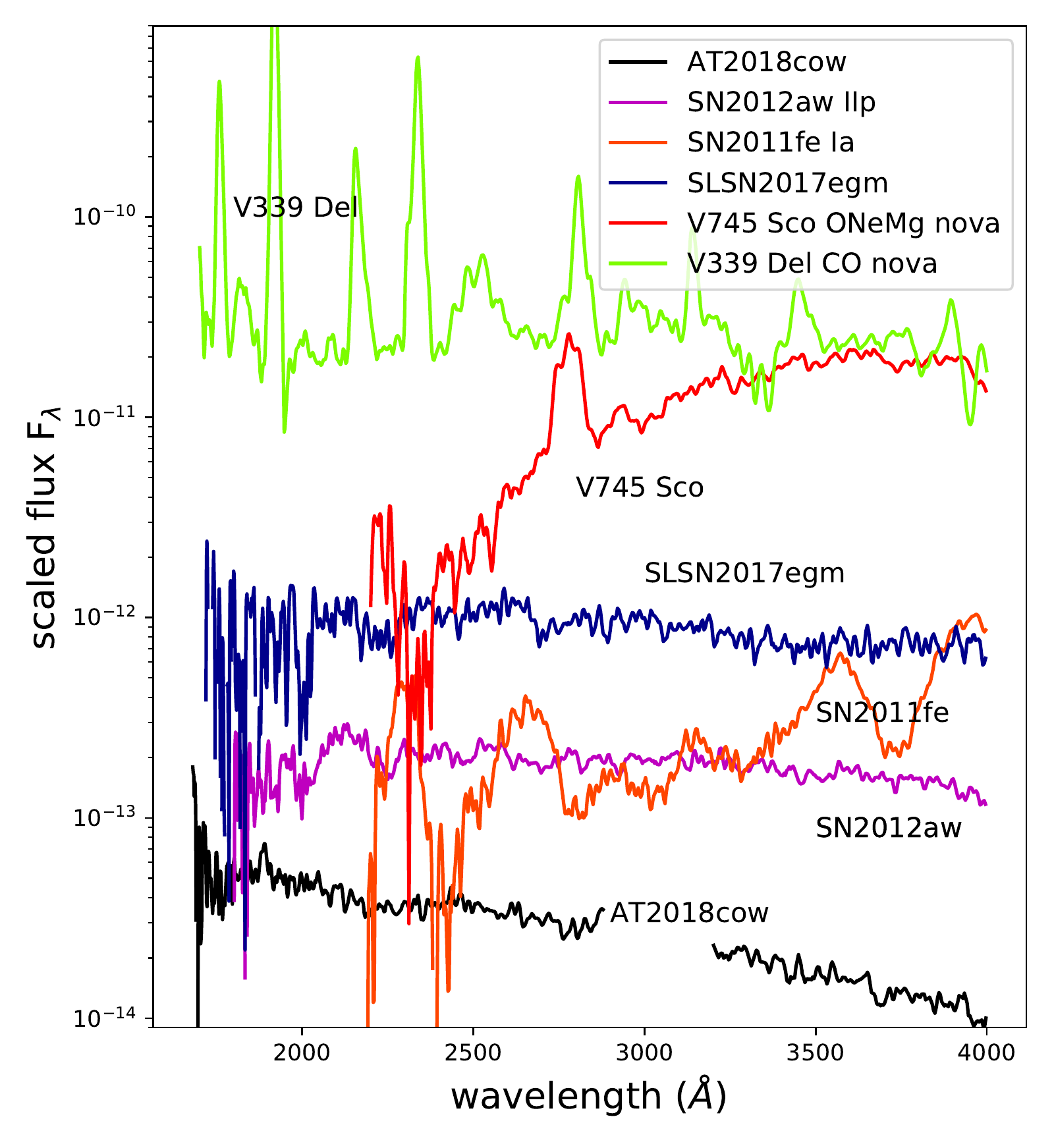}
\caption{UVOT grism spectra of various transients illustrate the difference with AT2018cow.}
\label{fig:compare2cow}
\end{figure}

In Fig.~\ref{fig:compare2cow} we compare \Swift UV spectra which shows the early (day 5) spectrum of AT2018cow stands out when compared to a SN\,Ia (SN2011fe at day 4), or a SN\,IIp (SN2012aw) at day 5; SN\,IIp are brighter in the UV than SN\,Ia. The SN spectra show broad absorptions which are mostly due to singly ionised metals. The recent superluminous SN2017egm at day 24 displays a rather flat spectrum in contrast, but is not as UV-bright as AT2018cow. 
A comparison to a CO-type and an ONeMg-type nova (V339 Del, V745 Sco, respectively) shows the strong UV emission lines from the expanding novae shell which are typically from enhanced abundances of C, N, O, Ne and Mg.    

The distance and optical magnitude imply an intrinsic brightness of the transient at maximum (on \Td+1.46\,d) ,  
 $L_{\rm bol} \approx  1.7\times 10^{44}\,{\rm erg\,s}^{-1}$, \citep{2018arXiv180705965P} 
  which is large for a SN, and excludes a kilonova type event because it is too bright. 

\begin{figure}
\includegraphics[width=\columnwidth]{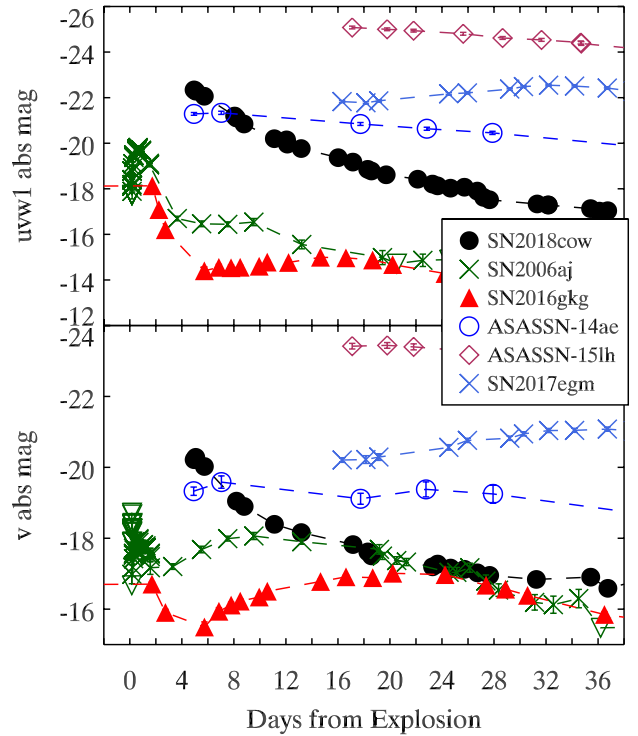}
\caption{The UV light curve of AT2018cow compared to those of the other UV-bright objects. }
\label{fig:SLSN}
\end{figure}

Fig.\,\ref{fig:SLSN} compares the absolute magnitudes of AT2018cow in the $uvw1$ and $v$ filters to other UV-bright objects: 
GRB060218/SN2006aj \citep{2006Natur.442.1008C}, 
the shock breakout and subsequent SN~2016gkg \citep{2017ApJ...837L...2A}, 
the superluminous supernova (SLSN)  2017egm \cite{2018ApJ...853...57B}, 
the  SLSN or tidal disruption event (TDE) ASASSN-15lh \cite{2016Sci...351..257D, 2016ApJ...828....3B, 2016NatAs...1E...2L}, 
and the TDE ASASSN-14ae \citep{2014MNRAS.445.3263H}. 

Photometry for all of these objects has been uniformly reduced using the \Swift Optical Ultraviolet Supernova Archive \citep[SOUSA]{2014Ap&SS.354...89B}; with distances and estimated explosion dates taken from the cited papers.

The high luminosity of AT2018cow strains models for SNe like SN2006aj and SN~2016gkg.  
Yet, the UV-optical emission is chromatic and presents a thermal spectrum, which is like a SN  \cite[see, e.g.][]{2014ApJ...780...21M}. 

Fast evolving luminous transients (FELT) which show a rapid rise and fast decay are proposed to arise when a supernova runs into external material thus lighting up a large area all at once \citep{2018NatAs...2..307R}. 
The FELT spectra show narrow emission lines from the re-ionised circum-stellar matter.  AT2018cow shows broad {\revv lines} in its spectra \citep{2018arXiv180705965P} and thus is not a FELT.

The X-ray flux ($0.3-10$\,keV) at \Td+3\,d is $10^{42}\,{\rm erg\,s}^{-1} \leq L_{\rm x} \leq 10^{43}\,{\rm erg\,s}^{-1}$ \citep{2018ATel11737....1R}.
Though the brightness is similar to a typical GRB, 
 we do not know any GRBs with $\gamma$-ray emission in the BAT ($14-195$\,keV) continuing for as long as 8 days.
However, the power law decay of the X-rays
with index 
$\alpha_1 = 0.9$ throughout,  
as well as the PL spectrum with $\beta = 1.6$ are 
similar to those associated with an off-axis jet  \citep{2009MNRAS.392..153D}.
The X-ray emission is therefore possibly due to a magnetic dominated jet of the kind we saw in Swift J1644 {\revv \citep{2012MNRAS.422.1625S}}, but it could also be that there is a more energetic GRB jet that we missed, while just seeing off-axis emission from that jet. 
The non-thermal emission from a jet may also explain the early detection in the radio  \citep{2018ATel11749....1D,2018ATel11774....1B} and point to a low CSM density.
Based on the X-Ray luminosity this could be a GRB; however, the luminosity ratio at day \Td+3\,d, $10 \leq L_{\rm opt}/L_{\rm x} \leq 100$, is large; in GRBs 11 hours after the trigger the ratio is less than 10 \citep{2014ARA&A..52...43B}. 
If it is a GRB, it is not a common type of GRB. 

In AGN X-ray flares are also commonly seen. 
However, the bursty flaring seen in the X-rays is closer to those of TDEs, e.g., \cite{2011ApJ...743..134K}. 
A comparison to a study of the peak X-ray flux in AGN and TDE also suggests the transient is a TDE \citep[their Fig.1]{2018ApJ...852...37A}. 
As the comparison of the light curves in Fig.\,\ref{fig:SLSN} shows, AT2018cow displays a faster evolution than the SLSN 2017egm, the SLSN or TDE ASASSN-15lh, or the TDE ASASSN-14ae, but its peak luminosity is in the same range. 

Tidal disruption is  a possible explanation of the observations. 
The observation of an emission source of radius $\approx 10^{15}$~cm as inferred from the blackbody fit (see Table~\ref{tab:uvot_bbmodel}) 
 which was formed within $\approx 3$\,d suggests a rapid, energetic event as would be the case for a tidal disruption by a black hole. 
The star being disrupted can therefore not be too large. 
Initially the observations showed only the spectral lines of He which suggest the small star could be a WD. 
Many WD have a magnetic field, and the formation of an energetic jet can be mediated by the remnants of the magnetic field after the outer parts of the WD have been tidally removed and also block the energy generated by accretion from escaping for at least part of the sphere. 
We explore in the following the jet and UV-optical debris resulting from the tidal disruption.

\subsection{The source of the $\gamma$ and X-rays}

\subsubsection{The jet associated with the TDE}

The power-law behaviour in the X- and $\gamma$-rays of the ``afterglow'' of AT2018cow brings to mind a jet. 
No {\revv high energy detector in orbit (Integral SPI-ACS and IBIS/Veto, Fermi GBM and LAT, INSIGHT HXMT/HE, ASTROSAT CZTI, MAXI GSC, Swift BAT) 
saw a prompt gamma-ray flash, with upper limits of a few times 10$^{-6}$ erg cm$^{-2}$ s$^{-1}$ for a short 0.1\,s bin and 
  of approximately $2 \times 10^{-7}$~erg~cm$^{-2}$~s$^{-1}$ 
  for a 10\,s bin size  at 10 keV to 100 MeV energies  \cite[e.g.,][]{2018ATel11843....1S,2018ATel11793....1D,2018ATel11799....1H,2018ATel11808....1K,2018ATel11809....1S,2018ATel11810....1S,2018Atel11782....1L} }, so it is not a common GRB jet. 
{\revv The upper limit corresponds to a few times $10^{49}$ erg s$^{-1}$ in the 1~keV to 10~MeV range, much higher than the brightness found later in the optical and X-rays, so there might have been undetected prompt $\gamma$-ray emission.}

We found in section~\ref{sec:BAT} that the gamma-ray emission  was not very energetic and of long duration. 
However, the slope of the light curve of $\alpha \approx 0.9$ prior to the break in the light curve at day 21, and the slope of the spectrum of $\beta = 1.6$ are indicative of a synchrotron dominated jet. 
We therefore investigated if the observed high-energy emission might be used to constrain an association with a GRB-like event. 
We can assume that the energy in the jet is imparted in the early stages of the disruption, and is likely to be of similar magnitude, or somewhat less, then the energy that propels out the massive debris which reached a radius of $5\times 10^{14}$\,cm in about a day. 
Using the photospheric density derived in section\,\ref{photosphere} we get {\revv a constraint for the energy in the jet and the kinetic energy in the debris ejecta: $E_{\rm jet} \leq 1.4\times 10^{50}$\,erg.} 

Based on the missing prompt $\gamma$-ray emission, if a GRB occurred, it must have been an ``off-axis'' event: if the opening angle of the ejecta that generate the unseen GRB jet emission is $\theta_{\rm j}$, the observer is placed at an angle $\theta_{\rm obs} > \theta_{\rm j}$. 
This way, the gamma-ray emission of the ultra-relativistic outflow was beamed away from the observer. 
The afterglow is instead visible because, in this phase, while the Lorentz factor $\gamma$ of the ejecta is lower than in the prompt emission from the core jet, the observer is within the cone of the beamed emission.

Analytical modeling \cite[see e.g.][]{2002ApJ...570L..61G}  and numerical modeling \cite[see e.g.][]{2010ApJ...722..235V}  indicates that, an early decay slope $\alpha \simeq 0.9$ like the one observed can be obtained if $\theta_{\rm obs} \simeq 1.25\;\! \theta_{\rm j}$. 
An afterglow seen off-axis is also considerably weaker than that on-axis; the above mentioned models indicate that, compared to the core jet, the flux decreases by a factor $\psi \simeq 0.1$. The hard spectral $\beta < 2$  index of the X-ray emission can be understood if $\nu_{\rm m} < \nu_{\rm x} < \nu_{\rm c}$, 
where $\nu_{\rm m}$ and $\nu_{\rm c}$ are the synchrotron peak and cooling frequency, respectively. 
In this configuration, $\beta = 1 + (p-1)/2$, where $p$ is the index of the power-law energy distribution of the electrons. In GRB afterglows $p < 3$ is usually seen, hence $\beta < 2$. In the following, we will assume $p=2.2$, i.e. 
the value of the decay slope given by deriving this parameter from $\beta$ in the case just described.

Given all the conditions above, the flux in the X-ray band $F$ at $10^{18}$~Hz (4.1\,keV) is 
\begin{eqnarray}
F & = &  5.1\times 10^{-14} \ 
\left( \frac{E_{\rm kin}}{10^{50}\;\!{\rm erg}} \right)^{1.3} 
\left({\frac{\epsilon_{\rm e}}{0.1} }\right)^{1.2} 
 \nonumber \\
& & \times\  \left({\frac{\epsilon_{\rm B}}{0.01} }\right)^{0.8} 
\left(\frac{ n }{{\rm cm}^{-3}}\right)^{0.5} \    
{\rm erg\, cm}^{-2} {\rm s}^{-1}\ .
\nonumber
\end{eqnarray}
\citep{2007ApJ...655..989Zhang}, 
{\revv where $E_{\rm kin}$, $\epsilon_{\rm e}$, $\epsilon_{\rm B}$ and $n$ 
  are the kinetic energy of the ejecta (assuming isotropy), 
  the fraction of energy in radiating electrons 
  and the fraction of energy in magnetic field, 
  and the number density of protons in the circum-expansion medium. 
With the measured flux (0.3 - 10~keV) 
  of $7.6 \times 10^{-12}$\,erg cm$^{-2}$ s$^{-1}$
  at \Td+21, and adding the correction factor for off-axis   
  emission $\psi$, the expression above gives  
\begin{eqnarray}
\left(\frac{\psi\, E_{\rm kin}}{10^{50}\;\!{\rm erg}}\right)^{1.3} 
\left(\frac{\epsilon_{\rm e}}{0.1}\right)^{1.2} 
\left(\frac{\epsilon_{\rm B}}{0.01}\right)^{0.8} 
\left(\frac{n}{{\rm cm}^{-3}}\right)^{0.5} \simeq 12 \ ,  
\nonumber
\end{eqnarray}
which can be easily satisfied 
  for typical values of parameters 
  found in GRB afterglows \cite[][]{2014ApJ...785...29S}.}

A consequence of collimated ejecta is that the observer should, at a given epoch, see a break simultaneously in the X-ray and in the optical band light curves. 
As the Lorentz factor of the ejecta decreases, more and more emitting surface becomes visible to the observer; 
 however, {\revv when  
  $\gamma \simeq (\theta_{\rm obs} + 2 \theta_{\rm j})^{-1}$,   
   no further emitting surface can be seen by the observer 
   \cite[][]{2010ApJ...722..235V}. } 
As a result, a steepening - ``jet break'' - of the afterglow light-curve ensues. Jet breaks should be a-chromatic, as in the case of AT2018cow. They are detected in GRB afterglows \cite[][]{2009ApJ...698...43R,2015ApJS..219....9W}. 
Post jet break decay slopes are $\alpha \simeq p$, which equates to $\alpha \simeq 2.2$ in our case. 
In our light-curves, we do not see such a fast decay up to $\simeq 21$~d. 
Modifying eq. 19 of \cite[][]{2009ApJ...698.1261Z} to take into account the off-axis position of the observer, 
{\revv  we derive a jet opening angle 
\begin{eqnarray} 
 \theta_j & = &  0.39 \left[ 
  \left(\frac{E_{\rm kin}}{10^{50}\;\!{\rm erg}} \right) 
  \left( \frac{{\rm cm}^{-3}}{n} \right)\right]^{-1/8} 
   {\rm rad} \ ,  \nonumber 
\end{eqnarray}
which depends only very weakly 
  on the kinetic energy in the ejecta 
  and number density of the circum-expansion media. 
Correcting $E_{\rm kin}$ for the jet opening angle 
  we find that sufficient energy 
  is available  to the jet 
  as it is less than the kinetic energy in the debris.}

\subsubsection{Emission from a shock running into the circum-system medium}
\label{sec:CSMshock}

For several days after the event a high velocity outflow, at a few percent of the speed of light, is inferred from the {\revv smooth broad features in the spectrum} \citep{2018arXiv180705965P,2018arXiv180800969Perley}.
The high-energy emission could be due to that initial outflow  shocking a pre-existing circum-system medium. 
However, the source of such a circum-system medium (CSM)  is not clear since mass loss from a WD is negligible and an origin from the black hole would require a previous interaction not too far in the past. 
The observations from \citet{2018arXiv180705965P,2018arXiv180800969Perley} show in the optical evidence of the high velocity outflow up to about day 9; we also notice continued evidence for $\gamma$-ray emission for the first 8 days, perhaps longer; and the X-ray emission shows a steady decline during that time, though with flares. 
The brightness of the high-energy emissions is such that a considerable CSM density is needed, however, we can place a limit to the CSM density using that the intrinsic column density from the fits to the X-ray spectra observations  $N_{\rm H} <  10^{20}$\,cm$^{-2}$, that the high velocity expansion is at $3\times 10^4$\,km\,s$^{-1}$ and that $\gamma$-rays take place over at least 8 days, so the CSM density must be lower than $5\times 10^4$\,cm$^{-3}$ {\revv \citep[see also][]{2018arXiv181010880Ho}. } 
Assuming optically thin radiative cooling \citep{1975A&A....40..355Hearn} and a temperature of 7 keV for the shocked gas, the total radiative loss would be less than $2.4\times 10^{33}$\,erg\,s$^{-1}$ which falls short of the observed X-ray luminosity by several orders of magnitude.

\subsection{AT2018cow as a tidal disruption event}    
\subsubsection{Condition for tidal disruption of a low-mass helium white dwarf} 

We propose that AT2018cow is probably caused by a TDE 
involving a mostly Helium white dwarf (hereafter He\,WD) or, alternatively, the remnant core of an evolved star, interacting with a black hole. 
The WD would be of spectral type DA with a thin Hydrogen atmosphere or of type DB, and likely is a field white dwarf.
The radius of tidal disruption of a star of radius $R_*$ and mass $M_*$ by a black hole of mass $M_{\rm bh}$  is roughly given by   
\begin{eqnarray} 
  R_{\rm t} & \approx & R_* \left( \frac{M_{\rm bh}}{M_*} \right)^{1/3}     \nonumber 
\end{eqnarray}  
\cite[see e.g.,][]{1975Natur.254..295Hills,2015JHEAp...7..148Komossa,2015JHEAp...7..158Lodato} 
The mass-radius relation of white dwarfs may be expressed as  
\begin{eqnarray} 
 \left(  \frac{R_{\rm wd}}{R_\odot}\right)^a \left(\frac{M_{\rm wd}}{M_\odot} \right)^b & =  &  \xi ~  f(M_{\rm wd},R_{\rm wd}) \ .    \nonumber 
\end{eqnarray}  
For low-mass He\,WDs, where general relativistic effects are unimportant,   
  $a = 3$, $b = 1$,  
  $\xi \approx 2.08 \times 10^{-6}$ and $f(M_{\rm wd},R_{\rm wd}) \approx 1$ 
\citep{2018GReGr..50...38C}. 
This implies that 
\begin{eqnarray} 
  R_{\rm t} & \approx &  R_\odot ~ \xi^{1/3} \left(\frac{M_{\rm bh}}{M_\odot} \right)^{1/3} \left(\frac{M_\odot}{M_{\rm wd}} \right)^{2/3}  \ . \nonumber 
\end{eqnarray}

The ratio of the radius of tidal disruption $R_{\rm t}$ to  
the Schwarzschild radius of the black hole $R_{\rm s}~ (= 2 GM_{\rm bh}/c^2)$ is given by   
\begin{eqnarray} 
    \frac{R_{\rm t}}{R_{\rm s}} & \approx  & \frac{\xi^{1/3}}{2} \left[ \frac{R_\odot c^2}{G M_\odot} \right] \left( \frac{M_\odot}{M_{\rm bh}}\right)^{2/3} \left( \frac{M_\odot}{M_{\rm wd}}\right)^{2/3} \nonumber \\ 
      & = & 3.01 \times 10^3 \left( \frac{M_\odot}{M_{\rm bh}}\right)^{2/3} \left( \frac{M_\odot}{M_{\rm wd}}\right)^{2/3} \ .  \nonumber  
\end{eqnarray} 
For a TDE to occur requires $R_{\rm t} > R_{\rm s}$.  
Thus, setting a lower limit of $0.1\;\!M_\odot$ to the mass of the He WD immediately constrains 
   the mass of the black hole to be $<1.3 \times 10^6\!\; M_\odot$.   

\subsubsection{Energetic considerations and constraints on the system parameters} 
\label{sec:BHmass}

While setting a lower limit to the WD mass gives the upper limit to the black-hole mass from the TDE criteria, 
 the luminosity produced in the event provides a means to constrain the minimum black hole mass.   
In order for the model to avoid self-contradiction,  
  the lower limit to the black hole mass as inferred from the observed luminosity 
  must not be larger than the upper limit to the black hole mass derived from the TDE criterion. 
Thus, the two will serve as independent assessments of the validity of the He\,WD TDE scenario,    

Without losing much generality, 
 we consider a spherical accretion of a neutral plasma 
  (the debris of the disrupted star) into the black hole.   
For a luminosity $L$ generated in the accretion process, 
  a radiative force $F_{\rm rad}$ will be generated and act on the charged particles in the inflowing plasma.  
The radiative force acting on an electron is simply   
 \begin{eqnarray} 
  F_{\rm rad} & = & \frac{ \sigma_{\rm Th}L}{4\pi r^2 c}  \  ,   \nonumber 
\end{eqnarray} 
 where $\sigma_{\rm Th}$ is the Thomson cross section 
 and $\bar \nu$ is the characteristic frequency of the photons.   
On the other hand, the gravitational force acting onto the plasma per electron is  
\begin{eqnarray}  
  F_{\rm g} & = & \frac{GM_{\rm bh}(x\;\!m_{\rm b}+m_{\rm e})}{r^2}   \ ,     \nonumber 
\end{eqnarray} 
where $m_{\rm b}$ is the mass of the baryons (which are the constituent protons and neutrons of the nuclei).  

Assuming a pure helium plasma (note though, that there can be a substantial amount of Hydrogen in a DA WD), $x\approx 2$. 
Equating $F_{\rm rad} = F_{\rm g}$ 
  gives the critical (Eddington) luminosity 
\begin{eqnarray} 
  L_{\rm Edd} & \approx & 2.56 \times 10^{38} \left( \frac{M_{\rm bh}}{M_\odot} \right) {\rm erg~s}^{-1}  \ . \nonumber  
\end{eqnarray} 
The corresponding Eddington mass accretion rate is given by 
  ${\dot m}_{\rm Edd} = L_{\rm Edd}/\lambda c^2$, 
  where the efficiency parameter $\lambda$ is of the
  order of $\sim 0.1$ for accreting black holes. 
Similarly, we may obtain the effective rate of mass accretion, the inflow 
  that powers the radiation, and that may be expressed as    
  ${\dot m}_{\rm in} = L_{\rm bol}/\lambda c^2$, 
  where $L_{\rm bol}$ is the bolometric luminosity of the radiation. 

Critical Eddington accretion and super-Eddington accretion 
  are generally accompanied by a strong radiatively driven mass outflow. 
Thus, we have   
  ${\dot m}_{\rm in}  =   {\dot m}_{\rm tot} -  {\dot m}_{\rm out}   
   = \eta ~{\dot m}_{\rm tot}$, 
  where $\eta$ is the fractional amount of inflow material 
  which contributes to the production of radiation 
  and ${\dot m}_{\rm tot}$ is the total mass that is  
  available for the accretion process in the tidal disruption event 

It is useful to define a parameter 
   $\zeta = {{\dot m}_{\rm tot}}/{{\dot m}_{\rm Edd}}$   
  to indicate how much the Eddington accretion limit is violated. 
With this parameter 
  we may express the bolometric luminosity of the accretion-outflow process as 
\begin{eqnarray} 
  L_{\rm bol} & = & L_{\rm Edd} \frac{{\dot m}_{\rm in}}{{\dot m}_{\rm Edd}}  \nonumber \\ 
     & = & 2.56 \times 10^{38}~ \eta\;\! \zeta \left( \frac{M_{\rm bh}}{M_\odot} \right) {\rm erg~s}^{-1}  \ .   \nonumber 
\end{eqnarray} 
It follows that  
\begin{eqnarray} 
 \left(\frac{M_{\rm bh}}{M_\odot} \right) & \approx & \frac{7.8\times 10^5}{\eta \;\! \zeta} 
    \left(\frac{  L_{\rm bol} }{2.0\times 10^{44}\;\! {\rm erg~s}^{-1}} \right) \ . \nonumber 
\end{eqnarray} 
for given $\eta$ and $L_{\rm bol}$, 
  $M_{\rm bh}(\zeta) > M_{\rm bh}(\zeta_{\rm max})$.  
The maximum degree of violation of the Eddington limit 
  in the subsequent accretion process after the WD disruption,  
  $\zeta_{\rm max}$, sets the lower mass limit of the black hole 
  that is allowed for the TDE. 
The remaining task now is to determine the parameter $\zeta$ 
  empirically using the observations.

\subsubsection{The scattering photosphere}
\label{photosphere}
The strong radiatively driven outflow 
  in the critical- or super-Eddington regime 
  will inevitably create a dense Thomson/Compton scattering photosphere. 
For a BH mass $\leq 10^6$\,M$_\odot$ the photosphere likely resides in the outflow 
  \citep[see e.g.][]{2009MNRAS.400.2070S}.  
Imposing an opacity $\tau_{\rm sc} \approx 1$ for a photosphere gives 
\begin{eqnarray}  
  n_{\rm e} & \approx &  \left( \sigma_{\rm Th} r_{\rm ph}\right)^{-1} \nonumber \\ 
    & = & 3.0 \times 10^9 
    \left(\frac{r_{\rm ph}}{5.0\times 10^{14}\;\!{\rm cm}} \right)^{-1} \;\!{\rm cm}^{-3} \ , \nonumber    
\end{eqnarray} 
  where $n_{\rm e}$ is the mean electron density in the photosphere 
  and $r_{\rm ph}$ is the characteristic photospheric radius. 
For a fully ionised helium plasma, the baryon (i.e. proton and neutron) 
  number density $n_{\rm b} = 2\;\! n_{\rm e}$.   
The total mass enclosed in the ionised helium scattering photosphere is 
\begin{eqnarray}  
  m(< r_{\rm ph}) & \sim &   \frac{4\pi}{3} {r_{\rm ph}}^3 n_{\rm b} m_{\rm b} \nonumber \\ 
    & \approx & 5.2\times 10^{30} \left( \frac{r_{\rm ph}}{5.0\times 10^{14}\;\!{\rm cm}} \right)^2\;\! {\rm g}  \ ,  \nonumber 
    \nonumber 
\end{eqnarray}  
  which is only a {\revv small} fraction of the mass of the He WD in disruption. 
The thermal energy contained in the scattering photosphere is roughly given by  
\begin{eqnarray} 
  E_{\rm th} (< r_{\rm ph}) & \approx & \frac{4\pi}{3}\;\! {r_{\rm ph}}^3  
    \left[\;\!\frac{3}{2} k_{\rm B} \left(n_{\rm e}T_{\rm e}+n_{\rm b} T_{\rm b} \right)\;\!\right]   \ . \nonumber
\end{eqnarray} 
Assuming thermal equilibrium between the baryons and the electrons, 
  $T_{\rm b} = T_{\rm e} = T$, and   
 \begin{eqnarray} 
  E_{\rm th} (< r_{\rm ph}) & \approx  &6\pi \;\! {r_{\rm ph}}^3 ( n_{\rm e} k_{\rm B} T ) \nonumber \\ 
    & = & 1.1 \times 10^{47} \left( \frac{r_{\rm ph}}{5.0\times 10^{14}\;\!{\rm cm}}   \right)^2  
      \left( \frac{k_{\rm B}T}{10~{\rm keV}} \right)\;\!  {\rm erg}   \ ,  \nonumber 
\end{eqnarray} 
  which is also a small fraction of the total energy produced in the TDE. 

The photospheric radius is the boundary at which 
  the outflow beyond it will become transparent. 
At the photospheric radius  the mass outflow rate is   
\begin{eqnarray} 
  {\dot m}_{\rm out} (r_{\rm ph}) & \sim &  4\pi \;\! {r_{\rm ph}}^2 \big[ \;\! n_{\rm b} m_{\rm b}\;\! v(r_{\rm ph}) \;\! \big] \nonumber \\ 
    & \approx & 6.3\times 10^{24}  \left( \frac{r_{\rm ph}}{5.0\times 10^{14}\;\!{\rm cm}}   \right)  \left( \frac{ v(r_{\rm ph})}{2000\;\!{\rm km~s}^{-1}}   \right) 
     \;\!{\rm g~s}^{-1} \ ,    \nonumber  
\end{eqnarray}  
  where $v(r_{\rm ph})$ is the outflow speed at the photospheric radius.  
Since this is a supersonic outflow, this mass will be lost to the system. 
Recall that $L_{\rm bol} = \lambda\;\! {\dot m}_{\rm in} c^2$ 
  and ${\dot m}_{\rm in} = \eta \;\!{\dot m}_{\rm tot}$. 
We then have 
\begin{eqnarray}  
  L_{\rm bol} & \approx & 5.7 \times 10^{44}  \times \left( \frac{\eta}{1-\eta}\right) \left(\frac{\lambda}{0.1}\right)  \nonumber \\
   & & \left( \frac{r_{\rm ph}}{5.0\times 10^{14}\;\!{\rm cm}}\right) 
   \left( \frac{ v(r_{\rm ph})}{2000\;\!{\rm km~s}^{-1}}   \right)     
    \;\!{\rm erg~s}^{-1} \ ,    \nonumber  
\end{eqnarray}  

\begin{figure}
\includegraphics[width=\columnwidth]{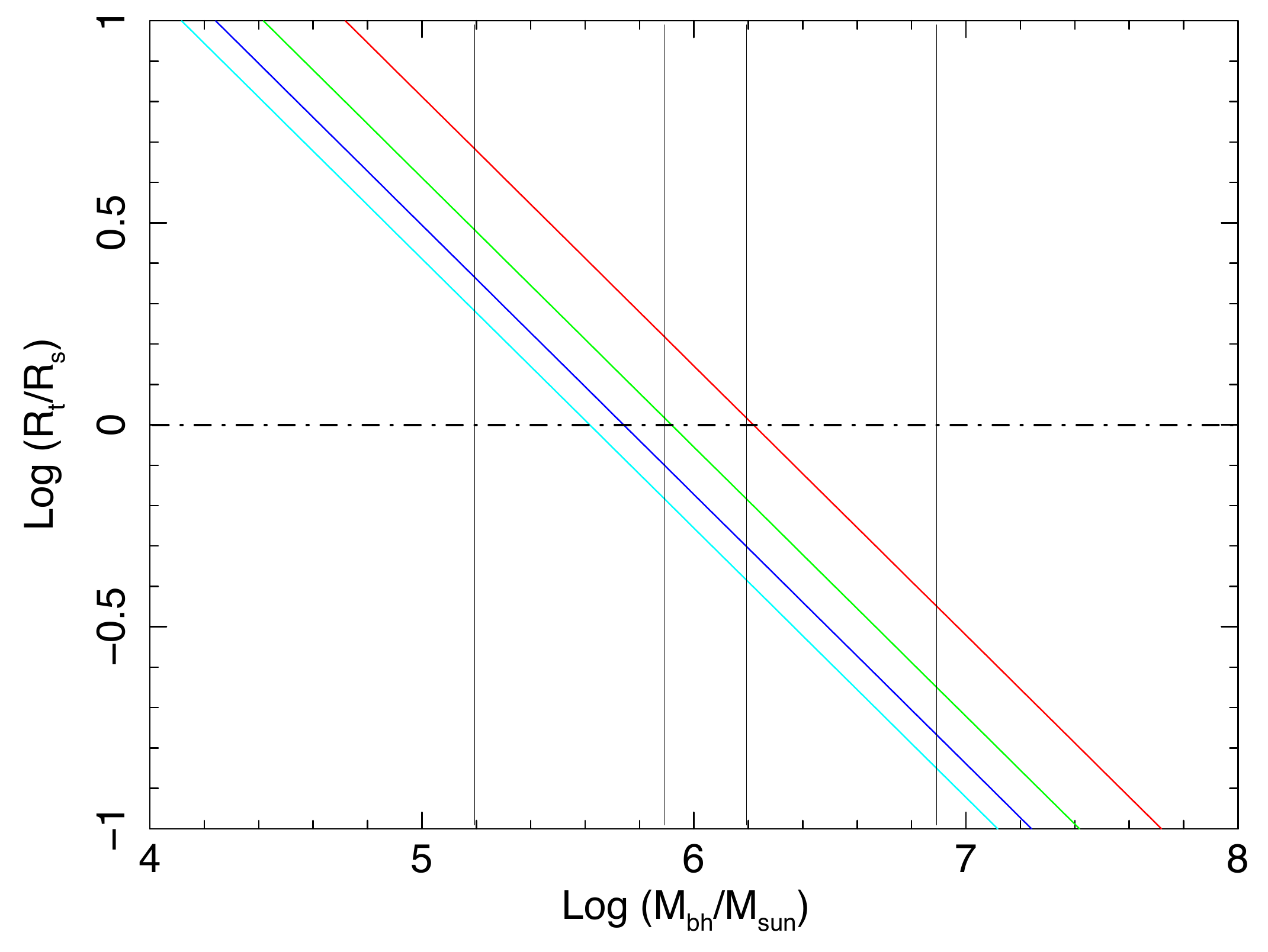}
\caption{
 Constraints to the systems parameters for a tidal disruption event  
   in a $\log\;\! (R_{\rm t}/R_{\rm s})$ vs $\log\;\! (M_{\rm bh}/M_\odot)$ plot, 
   where $M_{\rm bh}$ is the black hole mass and $R_{\rm s}$ is the Schwarzschild radius.  
 The inclined coloured lines correspond to the 
   tidal disruption radii for helium white dwarfs 
   of masses 0.1, 0.2, 03 and $0.4\;\! M_\odot$
   (from top to bottom respectively).   
 Tidal disruption is allowed when $R_{\rm t} > R_{\rm s}$, 
   and the critical condition where $R_{\rm t} = R_{\rm s}$
   is indicated by the dotted-dashed horizontal line.  
 The vertical lines indicate the lower bounds of the black hole masses  
   for the parameters $\eta \zeta = 5$, 1, 0.5 and 0.1 
   (from left to right respectively).  }
\label{fig:WD_TDE_model}
\end{figure}

The constraints on the system parameters based on the above considerations  
  can be seen in Fig.~\ref{fig:WD_TDE_model}. 
For a TDE to occur, the value of $\eta \zeta$ must be larger than 0.5 
  as restricted by the upper mass limit of the black hole 
  which is about $1.3 \times 10^6M_\odot$.  
Moreover, the black hole 
  would be more massive than $1.3 \times 10^5 M_\odot$ for $\eta \zeta < 5$, 
  i.e. the Eddington mass accretion limit is not too strongly violated.
More specifically, 
  if we set $\lambda = 0.1$, $r_{\rm ph} = 5\times 10^{14}\;\! {\rm cm}$ 
  and $v(r_{\rm ph}) \approx 2000\;\! {\rm km~s}^{-1}$,   
  then a bolometric luminosity of $2\times 10^{44}\;\!{\rm erg~s}^{-1}$ 
    will give $\eta \sim 0.26$. 
For a $10^6\;\!M_\odot$ black hole, 
  $\zeta \sim 3$ and ${\dot m}_{\rm in}/{\dot m}_{\rm Edd} \sim 0.78$;  
  whereas for a $5\times 10^5M_\odot$ accreting black hole, 
  $\zeta \sim 6$ and ${\dot m}_{\rm in}/{\dot m}_{\rm Edd} \sim 1.6$.  
These values are {\revv plaus}able for a black hole being practically forced-fed in a TDE.  

{\revv 
Note that the BH could be rotating. 
Some modification of the analysis in which we have adopted a Schwarzschild BH would be required 
  in order to take account of a smaller event horizon 
  for a Kerr black hole of the same mass. 
For instance, the tidal disruption radius could be 50\% smaller 
  for a prograde entry of the disurpted star but could be larger for a retrograde entry \cite[see e.g.][]{2017MNRAS.469.4483Tejeda}. 
Thus, 
  a black hole mass higher than the limit 
  set by the analysis of the WD TDE with a Schwarzschild BH 
  would be allowed if the BH is rotating. 
Moreover, relativistic effects would be non-negligible 
  when the WD penetrates to a distance 
  comparable with the BH gravitational radius 
  regardless of whether the BH is rapidly spinning or not. 
Nonetheless, the uncertainty 
  that this would introduce would be of a $\sim 10\%$  level \cite[see e.g.][]{2017MNRAS.469.4483Tejeda}. 
}

\subsubsection{Evolution of the photosphere and its atmosphere}

\begin{figure}
\includegraphics[width=\columnwidth]{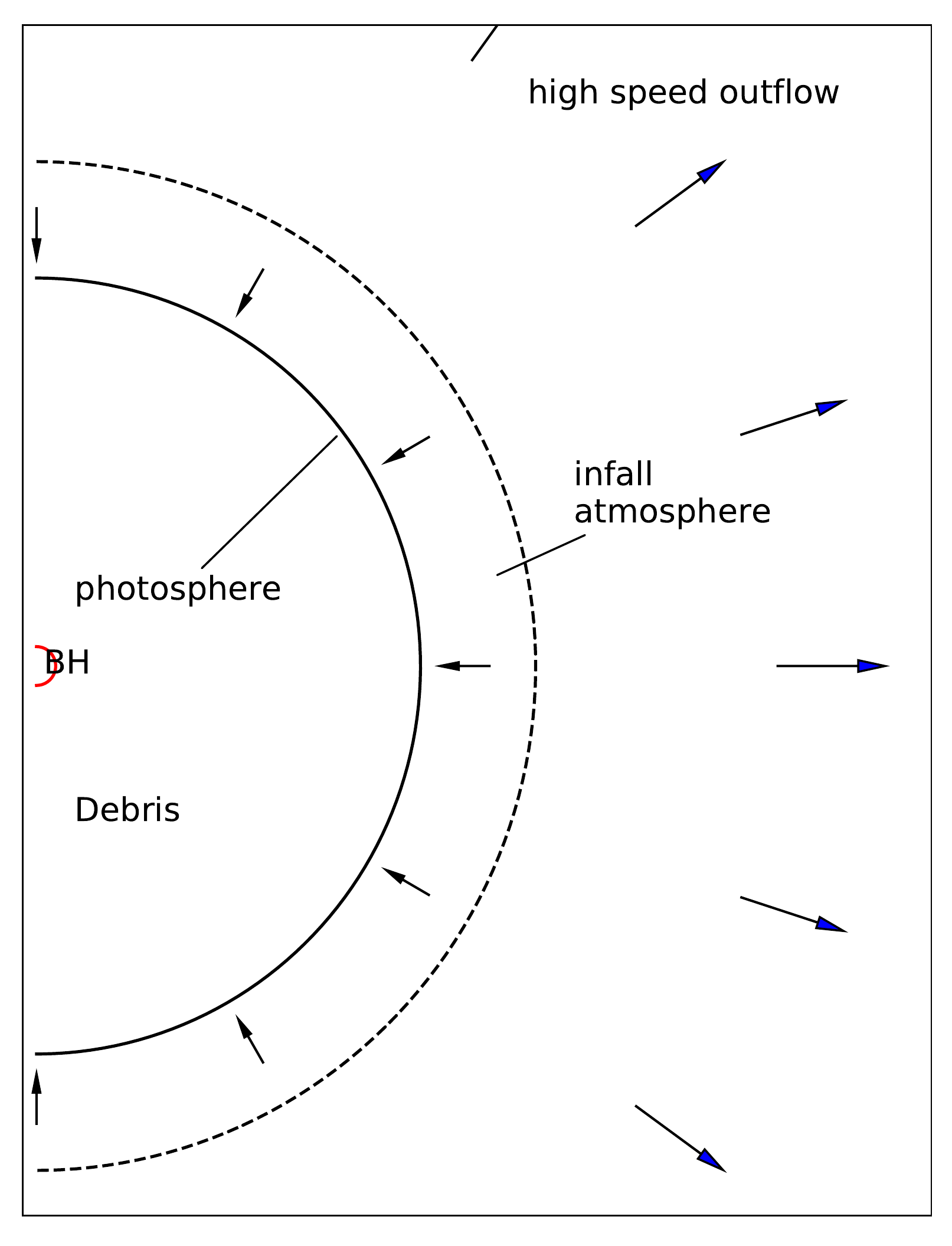}
\caption{A schematic illustration (not in scale) of the model of the disrupted WD debris around the BH of AT2018cow. 
The debris of the tidal disruption forms a photosphere, its associated enveloping atmosphere, and a high velocity outward expanding transient cocoon.
The line and continuum emission formation regions have been indicated as respectively, infall-atmosphere and photosphere. 
The atmosphere, which is optically thin, eventually falls back in while the photosphere cools and recedes inward and the atmosphere emits line profiles that loose their blue wings due to that infall.
The $\gamma$-ray and X-ray emitting jet is not included.   
  }
\label{fig:cartoon}
\end{figure}

The optical/infrared spectra of the source show a certain amount of excess emission above 
  a continuum, which has been well-fit by a blackbody spectrum, 
  and there is also clear evidence of distinctive emission lines in the later stages \citep{2018arXiv180800969Perley}. 
During the late stages (day \Td+11 and later) there is obvious evidence of He emission as indicated in the sequence of spectra 
  presented in \cite{2018arXiv180705965P} and also in 
 \citet{2018arXiv180800969Perley}. 
The later stages also show lines from the H Balmer series. 
Spectral lines due He-burning, i.e., C, N, and O, were however absent, as in our UV spectra. 
For an evolved star, the presence of a substantial amount of He, together with the lack of CNO elements would require that the star involved in the TDE cannot be a very massive star.  
Accepting the dominant presence of He in forming the spectrum, 
and interpreting the broad bumps seen in both the spectra shown in \cite{2018arXiv180705965P} 
and  \citet{2018arXiv180800969Perley} during the 8 day-long initial  stage as velocity broadened He I emission, 
a large outflow velocity reaching a substantial fraction of the speed of light would be required. 
We argue that these lines were probably emitted from a fast expanding shock heated cocoon produced by the TDE, see Fig.\,\ref{fig:cartoon}.
Such a high expansion velocity is in fact consistent with our proposed TDE induced scattering photosphere scenario,  
  as the photosphere would need to be inflated to a radius of $\sim 5\times 10^{14}\,{\rm cm}$ within 3\,d, which requires $V \sim 0.1c$.   
Note that however the expanding cocoon would detach from the scattering photosphere quickly 
and it would become depleted in density during its outward propagation. 
At the same time the thermalised photosphere would be maintained by fall-back; infall of the debris of the disrupted star providing the fuel by accretion into the black hole.

Kinetic energy will dominate the expanding cocoon energy balance   
while its ionised plasma emits the very broad He lines whose intensity drops when the density of the cocoon decreases with expansion.
As shown in section~\ref{sec:CSMshock} the shock of the interaction of the cocoon with the CSM is not sufficient to power the observed high-energy emission.
  
Note that the spectral evolution in the optical/infrared band observed by \citet{2018arXiv180800969Perley}  provides information about the thermal evolution of the scattering photosphere and its atmosphere. 
While the scattering photosphere produces the (black body) continuum, 
the lines originate from a surrounding lower-density atmosphere.   
The He\,I\,5876\,\AA~ line is not expected to be heavily contaminated by other lines, and here we use it to illustrate the line formation process and the evolution of the line profile in terms of the photospheric and atmospheric emission processes. 
We start with summarising the key features regarding the line strength development and the profile evolution from  the \citet{2018arXiv180800969Perley} observation. 
We ignore the broad bumps present in the early stage spectra {\revv and SEDs},  as they were formed in the rapidly expanding cocoon instead of being associated with the more stationary optical/infrared photosphere.  

The He\,I~5876~\AA\  emission line was not obvious in the spectra before day\,11. 
It began to emerge, with a symmetric broad profile centred at a frequency red-ward of the rest-frame line frequency. 
Though the line strength relative to the continuum increased, in reality both the continuum and the total line flux actually decrease quite substantially. 
The line starts showing asymmetry with its peak migrating blue-ward, toward the rest-frame line frequency while the blue wing is decreasing in intensity. 
By day\,33 and afterwards its profile becomes extremely asymmetric, peaking sharply almost exactly at the rest-frame line frequency, but only emission from the red wing is present. 
Almost no line flux remains in the frequencies blue-ward of the rest-frame line frequency. 
There is however no evidence of a P-Cygni absorption feature, an indicator of cooler outflow surrounding or within the line formation region,  \cite[see Fig. 4 in ][]{2018arXiv180800969Perley}. 
   
These line properties can be explained nicely with a simple two-zone model  in which a photosphere, which is optically thick to both the line and the continuum,  is enveloped by a lower density atmosphere, which is relatively opaque to the line emission  but much less so to the continuum emission.  
The continuum is being emitted from the photosphere whose boundary is defined by an optical depth $\tau_{\rm con} = 1$,  and the line is being generated in the atmosphere, mostly at an optical depth $\tau_{\rm line} = 1$  when the atmosphere is opaque to the line emission. 
For the special case when the photosphere and atmosphere are in local thermal equilibrium,  the intensities of the line and its neighbouring continuum will be the same, characterised by a Planck function at their equilibrium temperatures,   i.e. $T_{\rm ph} = T_{\rm at}$. 
It follows that $B_\nu(T_{\rm ph})\vert_{\rm con} = B_\nu(T_{\rm at})\vert_{\rm line}$, and hence no emission features but only a smooth thermal blackbody continuum will appear in the spectrum.  
     
Emission lines emerge only when the temperature of the line emitting atmosphere becomes higher than the thermal temperature of the continuum emitting photosphere. 
If the effective thermal temperature of the photosphere drops very substantially, because of a rapid cooling,  while the temperature of the lower-density atmosphere decreases at a slower rate, due to a less efficient cooling, the atmosphere would become hotter than the photosphere below.  
As a consequence, emission lines appear in the spectra together with the blackbody continuum from the photosphere. 

With this emission line formation mechanism in mind,  we now readily explain the extreme asymmetric profile of the He\,I~5876\,\AA\   line that had almost no emission blue-ward of the rest-frame line frequency and the overall reduction in the total line flux with time, by means of (1) the radial collapse of both of the photosphere and its atmosphere  and (2) the cooling of the both the photosphere and its atmosphere and the difference in their relative rates.

\begin{figure}
\includegraphics[width=\columnwidth]{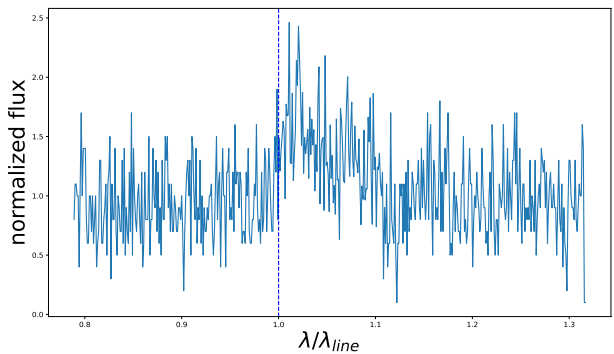}
\caption{Monte-Carlo simulation of the profile of the 
He\,I~5876\,\AA~ line formed in a geometrically thin atmosphere 
 (i.e. its velocity structure is assumed uniform). 
The atmosphere is optically thin to the continuum but not to the He\,I emission line.  
Note that the simulated line profile shows most emission above the continuum to the red wavelengths relative to the line rest wavelength (indicated with a vertical dashed line), while there is no contribution to the blue line wing. 
The reason is that the line is formed in infall back to the centre. 
This profile simulates well the characteristic 
shape of the observed line profile {\revv for days 37 and 44  in Fig. 4 of }
\protect\citet{2018arXiv180800969Perley}}.
\label{fig:line_profile}
\end{figure}

The peak emission of the line at the rest-frame frequency will be contributed 
  mostly by the limb region of the atmosphere where the line-of-sight projection of the infall velocity is essentially zero; 
  the reddest emission in the line wing will be contributed by the atmosphere above the central region of the photospheric disk, 
  where the line-of-sight projection of the infall velocity was the largest. 
Although the atmosphere is optically thick to the line it should be sufficiently optically thin that scattered photons from the whole visible atmosphere can escape. 
Otherwise, the line emission from the limb will  be suppressed by absorption of line photons due to the longer path length through the atmosphere.

In Fig.~\ref{fig:line_profile} we show that such an asymmetric line profile can be produced in simulations of line emission from a thin collapsing atmosphere of a finite thickness which is transparent to the continuum at frequencies near the  rest-frame line frequency.  
  
What remains to be explained now is why the line at first appeared to be reasonably symmetric with a peak to the red and later became extremely asymmetric and yet the line peak was at a frequency that was always red-ward of the rest-frame line frequency. 
We attribute the evolution of the line profile as being due to various radiative transfer effects, their convolution and combination.  
Among the effects, one is the line-of-sight attenuation in the atmosphere not sufficiently transparent to the continuum which is partly caused by the small difference in scale height between the regions of formation of continuum and line.   
The effects of the centre to limb variation in path length on the line formation also needs consideration.

To quantify the shift of the peak frequency and the development of the line profile with the change in the line and continuum opacities within the atmosphere and taking into consideration the thermal and dynamical structures of the photosphere and the atmosphere will require detailed radiative transfer calculations,  which are beyond the scope of the work.  
We therefore leave the line formation mechanism in a collapsing inhomogeneous plasma sphere to a separated future study. 
Nevertheless, from a simple geometrical consideration, 
   we may see that the optical depth on a curved atmospheric surface would vary according to $\tau \sim 1/\cos \Theta$  
   where $\Theta$ is angle between the line-of-sight and the normal to the atmospheric surface.      
Thus, the expectation is that the optical depth of the line would not show dramatic variations across the surface of a collapsing atmosphere as the one considered here. 
The symmetric profile of the He\,I lines observed in the source shortly after they emerge is expected therefore to 
evolve into an asymmetric profile on a timescale comparable to 
 the timescale on which the thermal coupling between the atmosphere and the photosphere becomes inefficient and the respective scale heights start to differ.

\subsubsection{AT2018cow compared to other TDE}

Tidal disruption events can be quite  energetic in their high energy emissions.
For example, the well studied {\em Swift}~J1644+57 transient \citep{2011Natur.476..421B} with large variability has peak isotropic X-ray luminosities exceeding  $10^{48}$\,erg\,s$^{-1}$, several orders of magnitude larger than the isotropic X-ray luminosity of AT2018cow, and also with a much harder spectrum. 

Spectra of the TDE PS1-10jh \citep{2012Natur.485..217G} were well-fitted by a galaxy model, a $3\times 10^4$\,K BB, and prominent broad He\,II emission lines on day \Td+22  indicative of velocities $9000\pm700\,{\rm km\,s}^{-1}$.
AT2018cow also has a BB and He lines, but
the  emission lines in PS1-10jh were well-defined.
The light curve of  PS1-10jh showed a slow brightening to a maximum around day 80 while the estimated peak luminosity is of the order of $10^{44}\,{\rm erg\,s}^{-1}$ \citep{2012Natur.485..217G}. 
Where the peak luminosity is similar, the light curve in PS1-10jh is quite different, indicative of an event with much larger intrinsic time scales.

Two UV studies of TDE for iPTF16fnl \citep[][]{2018MNRAS.473.1130Brown_uvTDE}, and for ASASSN-14li \citep{2016ApJ...818L..32Cenko_uvTDE} find their HST UV spectra are without the C\,III] line, but the N\,III]\,1750\,\AA~ line as well as higher excitation permitted lines of C and N are seen {\revv in the spectra that they discuss}. 
This was seen as an indication that the interacting star was around a solar mass main sequence star.

The absence of N\,III]\,1750\,\AA~ in our UV spectra is consistent with a He WD. 
A He-star, the stripped core of a one-time heavier star might seem possible also, but usually the core is not He through and through but envelops a CNO nucleus.

{\revv  
\subsubsection{On the absence of lines of certain elements}

One might wonder whether the absence of lines of C\,II, C\,III, N\,II, N\,III, O\,II in the UV are a result of an ionisation effect. 
We can discuss this in terms of a steady-state atmosphere, like a stellar photosphere, or in terms of a photosphere in a turbulent medium, like in nova ejecta. 
In the stellar context, ionisation of the CNO elements is either due to a high photospheric temperature, or due to non-radiative heating in a low density chromosphere above the photosphere. Emission lines result due to the rising excitation temperature above the photosphere, and lines of multiple stages of ionisation can be found in strengths proportional to the emission measure of the 
chromosphere, \cite[e.g.,][]{1982ESASP.176...83Jordan}.
The emission measure includes the elemental abundance and emission volume. 
A turbulent atmosphere also can have an embedded photosphere and emission lines of multiple ionisations can be found. 
An example would be ejecta in a nova. 
It is quite common to find a range of ionisations in astrophysical objects. 

In AT2018cow we interpret the spectrum as due to a photosphere at a temperature of 25,000 to 13,000\,K, in which ions of singly and doubly ionised CNO would be expected.
It is surrounded by a lower density medium which shows at late times He and H emission lines that also indicate excitation temperatures in the 10,000-30,000\,K range. 
As we argued before, line formation depends on the temperature of the line-forming region compared to that of the continuum. 
The evident He emission lines in the spectra prove that the excitation temperature in the lower density medium is sufficient to produce CNO lines provided that C, N, and O are abundant.

Models of the ionisation of the ejecta can be used to explore scenarios that might explain the absence of certain spectral lines at certain times from the TDE onset, like for example in \citet{2017ApJ...846..150Yang}.
In a hydrodynamical model of PS1-10jh \citet{2014ApJ...783...23G} describe how it is possible to explain the absence of certain spectral lines due to zones of ionisation in the ejecta. However, their model is for a case where the radiation pressure is negligible and that results in a completely different ionisation structure in the debris disk. Our approach for understanding AT2018cow starts with the observed luminosity, temperature, line-broadening and line-shape to explain the data in terms of a spherical cloud due to radiation pressure which results from a fast and energetic TDE. Assuming, like in \citet{2014ApJ...783...23G} that the radial extent was just due to hydrodynamics and limited by the escape speed would require a super-massive BH for AT2018cow. Therefore the ionisation structure of the AT2018cow debris cloud is likely very different.
}

\subsubsection{The host galaxy}

\begin{figure}
\includegraphics[width=\columnwidth]{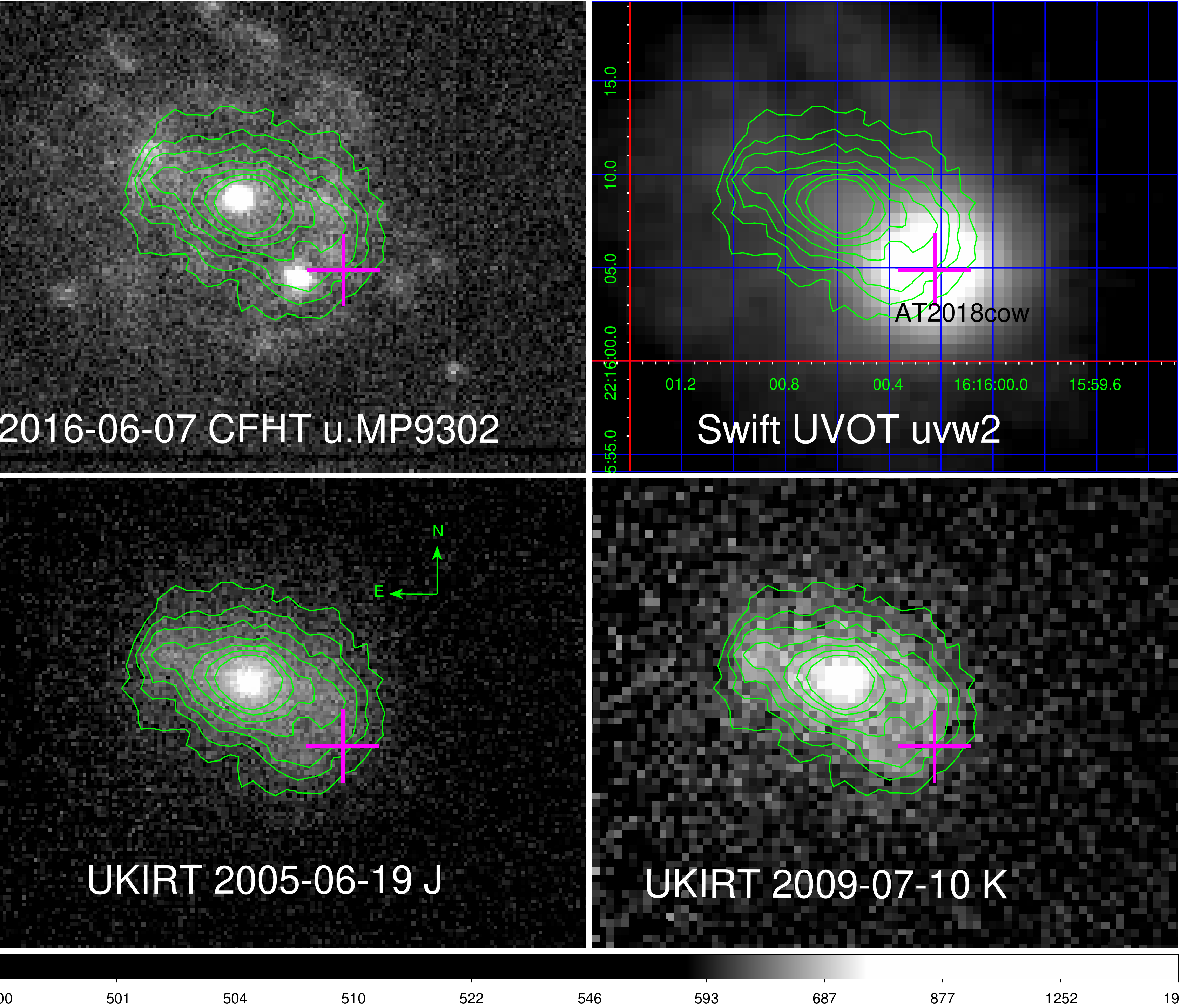}
\caption{Galaxy {\tt Z~137-068} in pre-explosion $u$ (CFHT), J, and K (UKIDSS) and after explosion {\em Swift} $uvw2$ image with AT2018cow position indicated by a cross. Contours are taken from the J-band image and superimposed on the others. The IR images in J and K show to the south-west extra emission that breaks the overall symmetry of the galaxy. The $u$ band image shows near the extra South-West emission a strong UV bright source. In the $u$ band  many small  small clusters of bright emission can be seen around the galaxy contour. }
\label{fig:galaxy}
\end{figure}

The nearby galaxy, see Fig.~\ref{fig:galaxy}, likely hosts an active galactic nucleus which might explain that the historic photometry at, e.g., $0.64-0.67\,\mu{\rm m}$ ranges by three orders of magnitude in brightness from 
  $6.8\times 10^{-5}-0.1$\,Jy \citep{2016A&A...595A...1G}, and \citep[POSS II:J]{2008AJ....136..735L}. 
The $u$ band image taken at the CFHT Megacam instrument \citep{1998SPIE.3355..614B} shows extended faint emission around {\tt Z~137-068} which look like H~II regions.
We retrieved the UKIDSS \citep{2007MNRAS.379.1599L} WFCAM \citep{2007A&A...467..777C} images in $J$ and $K$ taken by the UKIRT from the archive at ROE.
We also inspected GALEX FUV and NUV images, which due to the lower resolution show emission overlap between the galaxy bulge and the UV bright source. 
The UV bright source is clearly offset from the location of AT2018cow, however. 

The mass of the galaxy disk and bulge have been estimated as 
{\revv 
  $\log (M_{\rm disk}/M_\odot)  = 9.575$ 
  and $\log(M_{\rm bulge}/M_\odot) = 8.904$. Random uncertainties in the bulge mass are typically 0.15 dex, with additional systematic uncertainties of up to  60\% \citep{2014ApJS..210....3M}.}
The ratio of bulge mass to the mass of a central BH has been determined from known observations of central BH in galaxies. 
The recent \cite{2013ApJ...764..184M} correlation gives 
$M_{\rm bh} =1.8\times 10^6 M_\odot$ {\revv with uncertainties of about 0.3 dex in the BH mass. 
The uncertainties may be larger since the relations are typically derived for more massive galaxies. 
This }suggests that the central BH mass in the galaxy is slightly larger than the upper limit to the BH mass associated with a possible TDE in this transient{, \revv but it might be up to 10$^7 M_\odot$ when errors are considered}. 


The transient is offset from the nearby {\tt Z~137-068} galaxy by 6.0\arcsec which translates to $> 1.7$\,kpc  at the distance of 60.0\,Mpc. 
The extra emission near the location of the transient may be due to a a foreground object related to the excess nebulosity in the south-west area of the galaxy, since the narrow Ca~II lines in the spectrum of \cite{2018arXiv180705965P}  were found at a redshift of $z = 0.0139$ instead of $0.01414$ for {\tt Z~137-068}.
{\revv The difference between z=0.0139 and 0.01414 is 0.00024. The velocity difference is therefore 72 km\,s$^{-1}$. 
This is well within the rotation velocity of the galaxy, so it might be in {\tt Z~137-068}, or it could be a satellite galaxy of it, or it could be of order 1\,Mpc distant if $\Delta z$ is due to the Hubble flow.   
The positions of the late time spectral lines seem consistent with the transient association with 
{\tt Z~137-068}, but could also be consistent with the smaller redshift of 0.0139 since the broad and asymmetric  line profiles prevent an accurate redshift determination}.

The negligible  intrinsic $N_{\rm H}$ column density suggest that the transient is located in a low IGM density environment.  

One possibility is that the BH is in a globular cluster (GC) associated with the host galaxy. 
{\revv However, black hole masses inside GCs are estimated 
from $10^3 - 4\times 10^4 M_\odot$, i.e., intermediate mass BH (IMBH)  \citep{2009ApJ...697L..77R}.  
A BH mass of $4\times 10^4 M_\odot$ is below range of the mass we expect based on our analysis in Section~\ref{sec:BHmass}. } 
The old population of a GC could also have a higher WD proportion than the average galaxy.

In some TDE the presence of He was taken by some investigators as a sign of a WD rather than a MS or giant star impacting a black hole \citep[for example, Sw~J1644+57,][]{2011ApJ...743..134K}, but {\rev2 their observed time scales are much longer than AT2018cow. Therefore } the odds are that most of those are due to the interaction of a main sequence or giant star with a BH.

\subsection{Discussion}

We show that this event is unlike other transients in its combination of brightness, its evolutionary timescales, its spectrum, its abundances, and proceed to use the observed characteristics to constrain its properties in terms of a model of a Helium white dwarf TDE. 
Our observations point to a large debris cloud{ \revv which became larger than 33 AU within 1-3 days,} with embedded photosphere. Around that a lower density atmosphere showing infall, a transient high-velocity cocoon being blown off and also, possibly, a jet. 
The choice of a White Dwarf TDE rather than a main sequence star TDE is suggested by the size of the debris cloud being larger and its formation time shorter than would be expected in the disruption of a main sequence or giant branch star, whose larger radius and lower density mean that disruption takes place at larger scales, resulting in smaller debris scales. 
Fitting disruption of a main sequence star to these observations would therefore  require an extreme and unlikely accretion luminosity \citep{2018arXiv180800969Perley}.

In the first days of the event we see fewer flares in the X-rays than later. 
At the same time the blackbody luminosity light curve shows a flatter decay until day 6. 
Judging from those two facts and considering that both the jet and the photosphere are powered from the accretion, it may be that the initial {\revv accretion was nearly continuous and stable but became intermittent when the accretion rate decreased,} suggesting two accretion phases with the latter being more variable.

There is a possibility that in WD TDEs nuclear burning takes place \citep{2018MNRAS.477.3449K}. 
The UVOT UV spectra of novae who have ejecta of similar temperatures generally show strong emission lines of N\,II, N\,III, and C\,III, but in this case there is no sign of any emission from these lines in the UV spectra taken in the period of \Td+5$-$20\,d. 
This suggests that the abundance of these elements was not enhanced which is consistent 
  with the models in \cite{2018MNRAS.477.3449K} 
  likely for a 0.2\,M$_\odot$ or even less massive WD.

{\revv If the BH is rotating rapidly, and the WD approached the BH from outside the plane of its rotation, the WD debris is likely not limited to a thin disk. It may initially be a 3-D spiral shape as in \citet[][Section 5]{2017MNRAS.469.4483Tejeda}. 
It is likely that the debris will be ionised and turbulent due to the large luminosity from accretion and will generate a strong magnetic field \cite[see, e.g.,][]{2015PhR...601....1G}, which would lead to a further thickening of the debris due to turbulent pressure. 
Magnetic fields in the TDE of a rotating BH would also play a role in jet formation similar to the model described in \cite{2011ApJ...740L..27LeiZhang}.
The luminosity at optical maximum and thereafter is caused by the   large projected area of the debris atmosphere. Although that may mean that we  observe a disk-like cloud face-on, the odds are that we observe an extended, roughly spherical, debris cloud at a different angle, especially since the X-ray emission is consistent with an off-axis jet. 
A possible reason that the debris is extended in all directions by the time of optical maximum, is that relativistic precession, turbulent pressure,  magnetic field amplification, dissipation  and instability are causing expansion of geometrically thin debris.
We are not aware of numerical models of TDEs that include also the physics of magnetic field generation, current systems, and their evolution in a relativistic environment which might be the missing element in explaining why AT2018cow developed so rapidly and why the debris cloud would be extended. 
}

Our analysis of the BAT $\gamma$-ray and XRT X-ray emission could be explained by a jet associated with the TDE.
The observed bursty peaks in the X-ray luminosity are also seen in other TDE as would be expected from an unsteady accretion process driving a variable jet. 
However, the total column density to the X-ray source is low, and the spectrum does not notably change during the bursts.
{\revv The low column density suggests that the line of sight to the source of the X-ray emission is not obscured by the tidal debris. 
In such circumstances the variability in the X-rays might directly probe the accretion history onto the BH.}

The initial optical polarisation seen day \Td+4.6, 4.7, see \cite{2018ATel11789....1S}, is possibly due  to a slightly non-spherical shape of the line emission from the cocoon, and its disappearance is likely linked to the evolution of the cocoon. 

{\revv Apart from WDs, we consider whether some other low mass star might be the cause of this TDE and have a short time-scale of interaction due to their low mass. 
Very low mass ($M < 0.6\;\! M_{\odot}$) main sequence stars are not expected to have burned more than 70\% of their hydrogen \cite[e.g.,][]{1979ApJS...40..733M} by the end of their evolution.  The large amount of H in the tidal disruptions of a very low mass main sequence star would be at odds with the relative strength of the He lines in the spectrum and initial absence of similarly broadened H lines. 
The  He core from a stripped late type star might be a possible candidate. He-cores can be formed by mass transfer in binaries, but observationally they would appear as He WDs. 
Therefore the most likely candidate for  the star in this tidal disruption is a WD.
}
The proposed nature of this transient as being possibly a magnetar \citep{2018arXiv180705965P,2018arXiv181010880Ho}, a SN \citep{2018arXiv180706369R}, or a TDE of a solar mass star \citep{2018arXiv180800969Perley} are either unlikely or do not explain as many observations as our model of a WD TDE. 

By contrast, a low mass WD (e.g. DB or even a DA WD) 
is considered a good candidate for AT2018cow, 
as they consist mostly of He.  
WD are common; they comprise of about 12\% of the Galactic stellar population. 
The ``mean'' WD mass is about $0.6\;\!M_\odot$. 
The Gaia DR2 data has showed about 5\% of WD in the sample have mass less than 
 $0.3\;\!M_\odot$  (Gentille Fusillo, personal communication). 
(For the Gaia DR2 WD catalogue, see \citet[][]{2018arXiv180703315G}.)  
A recent study  of WDs within 20\,pc of the Sun  \citep[][]{2018MNRAS.480.3942H} gave a less biased, but more uncertain, estimate  that 2 out of 137 WDs have mass $< 0.3 M_\odot$. 
About 0.5\% of the galactic stellar population  would therefore be in the low mass WDs. 
Thus, assuming that a similar percentage holds in the host galaxy, the chance encounter of a low-mass WD by a massive black hole  would not be an extremely rare event in comparison with other stellar types.  
{\rev2 However, that does not take into account the effects of the change of loss cone radius with stellar density \cite[e.g.][]{2014ApJ...794....9Macleod}. Such considerations are however difficult to make since the environment of the 
event is not known because it is not in the centre of a galaxy.} 

We propose a simple model that explains the observations of the emergent line profiles and show that infall of the lower density atmosphere surrounding the shrinking photosphere can explain the line profile shape and evolution. 
{\revv The presence of infall as inferred  from the line profiles, may be due to the infall being confined to closed magnetic field. In the Sun chromospheric and transition region lines (formed at temperatures of 7000-50000\,K) show on average a redshift \citep{1993ApJ...408..735PaalBrekke} which is due to the higher density in filamentary structures. A high velocity outflow may in the presence of a magnetic field co-exist with infall into the photosphere. }  
Detailed modeling of the hydrodynamics and thermodynamics,  atomic physics and radiative transfer is needed to work out this simple model in more detail.



\section{conclusion}

The {\em Swift} data of AT2018cow, supported by other studies and reports suggests that possibly this was 
the tidal disruption of a He WD on a relatively small non-stellar mass black hole, 
  resulting in a large, hot, ionised debris cloud emitting a thermal spectrum with weak, broad line emission. 
A jet may also be associated with the event and emit a non-thermal X-ray spectrum; it is not as luminous as those seen in long GRBs.  
The X-ray component due to optically thick Compton scattering on the hot debris cloud is negligible, and X-ray emission from a shock caused when a high velocity cocoon encounters the circum-system medium is estimated to be non-detectable. 

We explain the multi-wavelength temporal behaviours of the source with a WD-TDE model, and the observations give a constraint for the WD mass to be $\approx 0.1 - 0.4\,M_\odot$ and the BH mass to be $\sim 1.3 \times 10^5 - 1.3 \times 10^6\,M_\odot$. 
The model also predicts the total accreted rate of mass accretion is $\approx 8.5\times 10^{24}{\rm g\,s}^{-1}$, see section~\ref{photosphere}, while we can use $\zeta \sim 3-6$ to also derive the ejected mass loss rate is $\approx 6.3\times 10^{24}\,{\rm g\,s}^{-1}$.
The model is consistent with the observed bursty peaks in the X-ray luminosity which are also seen in other TDE as would be expected from an unsteady accretion process.

\section*{Acknowledgements}
We acknowledge the efforts of the Swift planners.
Swift and NuSTAR Data were retrieved from the Swift and NuSTAR archive at HEASARC/GSFC, and from the UK Swift Science Data Centre. 
We also used the CFHT archive hosted at the Canadian Astronomy Data Centre operated by the National Research Council of Canada with the support of the Canadian Space Agency, and the WFCAM UKIRT data from the UKIDSSDR10PLUS data release from the WFCAM archive at the Royal Observatory Edinburgh.
This work has been supported by the UK Space Agency under grant ST/P002323/1 and the UK Science and Technology Facilities Council under grant ST/N00811/1. 
QH is supported by a UCL MSSL Summer Research Studentship.
SRO gratefully acknowledges the support of the Leverhulme Trust Early Career Fellowship.
SC acknowledges the support of under ASI-INAF contract I/004/11/1.
We benefited from a very useful review by the referee, and thank him/her for their suggestions.

\bibliographystyle{mnras}
\bibliography{at2018cow} 

\appendix{A}
\section{Data tables}
\begin{table}[]
\caption{{\em Swift} UVOT photometry (corrected for galaxy)}
\label{tab:uvot_photometry}
\begin{tabular}{llllr}
\hline
time & MJD & magnitude & magnitude & filter \\
(d since \Td) & (d) & (AB) & error & \\
\hline
3.062 & 58288.503 & 13.552 & 0.043 & U \\
3.062 & 58288.503 & 13.571 & 0.047 & UVW1 \\
3.063 & 58288.504 & 13.852 & 0.04 & B \\
3.064 & 58288.505 & 13.568 & 0.054 & UVW2 \\
3.065 & 58288.506 & 14.007 & 0.041 & V \\
3.066 & 58288.507 & 13.584 & 0.047 & UVM2 \\
3.782 & 58289.223 & 13.82 & 0.058 & UVW1 \\
3.784 & 58289.225 & 13.943 & 0.043 & U \\
3.785 & 58289.226 & 14.152 & 0.041 & B \\
3.789 & 58289.23 & 14.187 & 0.042 & V \\
3.792 & 58289.233 & 13.846 & 0.061 & UVM2 \\
5.063 & 58290.504 & 14.522 & 0.072 & UVW2 \\
5.155 & 58290.596 & 14.537 & 0.049 & UVM2 \\
5.252 & 58290.693 & 14.532 & 0.088 & UVW1 \\
5.254 & 58290.695 & 14.508 & 0.044 & U \\
5.256 & 58290.697 & 14.903 & 0.044 & B \\
5.259 & 58290.7 & 14.862 & 0.05 & V \\
6.251 & 58291.692 & 14.735 & 0.089 & UVW1 \\
6.253 & 58291.694 & 14.792 & 0.045 & U \\
6.255 & 58291.696 & 15.173 & 0.046 & B \\
6.256 & 58291.697 & 14.987 & 0.1 & UVW2 \\
6.259 & 58291.7 & 15.178 & 0.055 & V \\
6.264 & 58291.705 & 14.86 & 0.059 & UVM2 \\
6.478 & 58291.919 & 14.837 & 0.054 & UVW1 \\
6.74 & 58292.181 & 14.911 & 0.037 & U \\
6.747 & 58292.188 & 15.238 & 0.066 & UVW2 \\
6.842 & 58292.283 & 15.323 & 0.048 & B \\
6.845 & 58292.286 & 15.343 & 0.058 & V \\
6.85 & 58292.291 & 15.316 & 0.057 & UVM2 \\
8.37 & 58293.811 & 15.406 & 0.059 & UVW1 \\
8.371 & 58293.812 & 15.587 & 0.055 & UVW2 \\
8.372 & 58293.813 & 15.344 & 0.054 & U \\
8.373 & 58293.814 & 15.639 & 0.054 & B \\
9.143 & 58294.584 & 15.946 & 0.05 & UVW2 \\
9.169 & 58294.61 & 15.77 & 0.061 & UVW1 \\
9.17 & 58294.611 & 15.557 & 0.056 & U \\
9.171 & 58294.612 & 15.853 & 0.058 & B \\
9.172 & 58294.613 & 15.858 & 0.05 & UVM2 \\
9.177 & 58294.618 & 15.919 & 0.077 & V \\
10.138 & 58295.579 & 15.92 & 0.044 & UVW1 \\
10.145 & 58295.586 & 16.267 & 0.061 & UVW2 \\
10.172 & 58295.613 & 15.714 & 0.062 & U \\
10.172 & 58295.613 & 15.851 & 0.066 & B \\
10.175 & 58295.616 & 16.101 & 0.104 & V \\
10.177 & 58295.618 & 15.922 & 0.066 & UVM2 \\
11.166 & 58296.607 & 15.894 & 0.036 & U \\
11.175 & 58296.616 & 16.448 & 0.06 & UVW2 \\
11.234 & 58296.675 & 16.131 & 0.064 & UVW1 \\
11.237 & 58296.678 & 16.082 & 0.063 & B \\
11.242 & 58296.683 & 16.191 & 0.09 & V \\
11.244 & 58296.685 & 16.198 & 0.065 & UVM2 \\
12.224 & 58297.665 & 16.582 & 0.051 & UVW2 \\
12.658 & 58298.099 & 16.271 & 0.048 & UVW1 \\
12.859 & 58298.3 & 16.516 & 0.047 & UVM2 \\
12.863 & 58298.304 & 16.626 & 0.06 & UVW2 \\
\hline
\end{tabular}
\end{table}

\begin{table}[]
\contcaption{}
\begin{tabular}{llllr}
\hline
time & MJD & magnitude & magnitude & filter \\
(d since \Td) & (d) & (AB) & error & \\
\hline
12.954 & 58298.395 & 15.983 & 0.057 & U \\
12.956 & 58298.397 & 16.198 & 0.061 & B \\
12.963 & 58298.404 & 16.419 & 0.09 & V \\
14.18 & 58299.621 & 16.554 & 0.047 & UVW1 \\
14.283 & 58299.724 & 16.171 & 0.059 & U \\
14.284 & 58299.725 & 16.314 & 0.063 & B \\
14.287 & 58299.728 & 16.919 & 0.076 & UVW2 \\
14.291 & 58299.732 & 16.417 & 0.091 & V \\
14.293 & 58299.734 & 16.784 & 0.067 & UVM2 \\
15.158 & 58300.599 & 17.184 & 0.089 & UVW2 \\
15.176 & 58300.617 & 16.389 & 0.039 & U \\
15.21 & 58300.651 & 16.739 & 0.071 & UVW1 \\
15.213 & 58300.654 & 16.535 & 0.073 & B \\
15.217 & 58300.658 & 16.654 & 0.115 & V \\
15.219 & 58300.66 & 16.987 & 0.072 & UVM2 \\
16.437 & 58301.878 & 17.451 & 0.045 & UVW2 \\
16.508 & 58301.949 & 17.127 & 0.056 & UVW1 \\
16.51 & 58301.951 & 16.618 & 0.047 & U \\
16.511 & 58301.952 & 16.702 & 0.055 & B \\
16.515 & 58301.956 & 17.002 & 0.099 & V \\
16.516 & 58301.957 & 17.336 & 0.048 & UVM2 \\
18.036 & 58303.477 & 17.394 & 0.058 & UVW1 \\
18.082 & 58303.523 & 17.934 & 0.076 & UVW2 \\
18.343 & 58303.784 & 17.005 & 0.087 & B \\
18.348 & 58303.789 & 17.287 & 0.164 & V \\
18.35 & 58303.791 & 17.722 & 0.086 & UVM2 \\
18.635 & 58304.076 & 16.95 & 0.044 & U \\
18.844 & 58304.285 & 17.932 & 0.085 & UVW2 \\
19.962 & 58305.403 & 17.947 & 0.06 & UVW2 \\
20.195 & 58305.636 & 17.582 & 0.089 & UVW1 \\
20.196 & 58305.637 & 17.042 & 0.078 & U \\
20.197 & 58305.638 & 17.014 & 0.09 & B \\
21.259 & 58306.7 & 18.057 & 0.06 & UVM2 \\
21.265 & 58306.706 & 18.232 & 0.069 & UVW2 \\
21.526 & 58306.967 & 17.262 & 0.062 & U \\
21.527 & 58306.968 & 17.362 & 0.081 & B \\
21.532 & 58306.973 & 17.543 & 0.151 & V \\
21.827 & 58307.268 & 17.935 & 0.068 & UVW1 \\
22.496 & 58307.937 & 18.469 & 0.077 & UVW2 \\
23.249 & 58308.69 & 18.021 & 0.084 & UVW1 \\
23.252 & 58308.693 & 17.427 & 0.084 & B \\
23.257 & 58308.698 & 17.664 & 0.164 & V \\
23.259 & 58308.7 & 18.273 & 0.073 & UVM2 \\
23.553 & 58308.994 & 17.487 & 0.05 & U \\
24.284 & 58309.725 & 18.597 & 0.093 & UVW2 \\
25.245 & 58310.686 & 18.522 & 0.105 & UVW1 \\
25.246 & 58310.687 & 17.751 & 0.078 & U \\
25.247 & 58310.688 & 17.8 & 0.108 & B \\
25.252 & 58310.693 & 18.069 & 0.227 & V \\
25.254 & 58310.695 & 18.667 & 0.073 & UVM2 \\
26.35 & 58311.791 & 19.032 & 0.11 & UVW2 \\
26.679 & 58312.12 & 18.684 & 0.119 & UVW1 \\
26.68 & 58312.121 & 17.996 & 0.08 & U \\
26.681 & 58312.122 & 17.931 & 0.109 & B \\
26.684 & 58312.125 & 18.375 & 0.27 & V \\
26.686 & 58312.127 & 18.845 & 0.084 & UVM2 \\
27.045 & 58312.486 & 19.08 & 0.158 & UVW2 \\
29.311 & 58314.752 & 18.184 & 0.194 & B \\
29.313 & 58314.754 & 19.264 & 0.169 & UVW2 \\
29.315 & 58314.756 & 18.532 & 0.467 & V \\
29.316 & 58314.757 & 19.331 & 0.198 & UVM2 \\
29.379 & 58314.82 & 18.24 & 0.124 & U \\
\hline
\end{tabular}
\end{table}

\begin{table}[]
\contcaption{}
\begin{tabular}{llllr}
\hline
time & MJD & magnitude & magnitude & filter \\
(d since \Td) & (d) & (AB) & error & \\
\hline
30.653 & 58316.094 & 19.061 & 0.146 & UVW1 \\
31.085 & 58316.526 & 18.384 & 0.129 & U \\
31.087 & 58316.528 & 18.152 & 0.155 & B \\
31.09 & 58316.531 & 19.39 & 0.166 & UVW2 \\
31.094 & 58316.535 & 18.539 & 0.37 & V \\
31.097 & 58316.538 & 19.38 & 0.151 & UVM2 \\
34.144 & 58319.585 & 19.541 & 0.204 & UVW1 \\
34.146 & 58319.587 & 18.856 & 0.139 & U \\
34.147 & 58319.588 & 18.498 & 0.164 & B \\
34.151 & 58319.592 & 19.678 & 0.18 & UVW2 \\
34.154 & 58319.595 & 18.795 & 0.368 & V \\
34.157 & 58319.598 & 19.55 & 0.122 & UVM2 \\
37.625 & 58323.066 & 19.661 & 0.257 & UVW1 \\
37.627 & 58323.068 & 19.146 & 0.181 & U \\
37.628 & 58323.069 & 18.511 & 0.173 & B \\
37.631 & 58323.072 & 19.824 & 0.206 & UVW2 \\
37.634 & 58323.075 & 19.173 & 0.537 & V \\
37.637 & 58323.078 & 19.592 & 0.133 & UVM2 \\
39.318 & 58324.759 & 19.944 & 0.317 & UVW1 \\
39.32 & 58324.761 & 19.512 & 0.233 & U \\
39.321 & 58324.762 & 19.308 & 0.322 & B \\
39.324 & 58324.765 & 20.049 & 0.24 & UVW2 \\
39.329 & 58324.77 & 20.142 & 0.175 & UVM2 \\
41.22 & 58326.661 & 19.395 & 0.214 & U \\
41.221 & 58326.662 & 19.097 & 0.271 & B \\
41.225 & 58326.666 & 20.107 & 0.257 & UVW2 \\
41.23 & 58326.671 & 20.219 & 0.197 & UVM2 \\
41.714 & 58327.155 & 20.358 & 0.458 & UVW1 \\
44.6 & 58330.041 & 19.868 & 0.317 & U \\
44.602 & 58330.043 & 18.987 & 0.247 & B \\
44.605 & 58330.046 & 20.322 & 0.307 & UVW2 \\
44.611 & 58330.052 & 20.446 & 0.221 & UVM2 \\
47.226 & 58332.667 & 20.037 & 0.349 & U \\
47.227 & 58332.668 & 19.456 & 0.357 & B \\
47.229 & 58332.67 & 20.532 & 0.359 & UVW2 \\
47.234 & 58332.675 & 20.256 & 0.179 & UVM2 \\
50.911 & 58336.352 & 19.203 & 0.283 & B \\
50.913 & 58336.354 & 20.669 & 0.401 & UVW2 \\
50.917 & 58336.358 & 20.708 & 0.249 & UVM2 \\
53.465 & 58338.906 & 19.546 & 0.456 & B \\
53.466 & 58338.907 & 20.83 & 0.489 & UVW2 \\
53.468 & 58338.909 & 20.747 & 0.323 & UVM2 \\
56.588 & 58342.029 & 19.758 & 0.509 & B \\
56.594 & 58342.035 & 21.248 & 0.419 & UVM2 \\
60.141 & 58345.582 & 21.147 & 0.379 & UVM2 \\
69.481 & 58354.922 & 21.72 & 0.47 & UVM2 \\
\hline
\end{tabular}
\end{table}

\bsp	
\label{lastpage}
\end{document}